\documentclass[twocolumn,secnumarabic, amssymb,  amsmath,tightenlines,showpacs,superscriptaddress,preprintnumbers, aps]{revtex4-1}
\usepackage{amsfonts,amsbsy,graphicx,bm,color,txfonts,dcolumn,times,bbm}

\usepackage{graphicx}
\usepackage{dcolumn}
\usepackage{bm}
\usepackage{morefloats}
\usepackage{float}
\usepackage{amsfonts}
\usepackage{amsbsy}
\usepackage[colorlinks=true,linkcolor=blue,filecolor=blue,menucolor=yellow,urlcolor=blue,citecolor=blue,anchorcolor=blue]{hyperref}
\usepackage{times} 
\usepackage{color}
\usepackage{amssymb}

\usepackage{epstopdf}

\begin{document}


\title{Angular dependence of superconductivity in superconductor / spin valve heterostructures}

\author{Alejandro A. Jara}
\author{Christopher Safranski}
\author{Ilya N. Krivorotov}
\affiliation{Department of Physics and Astronomy, University of California, Irvine, California 92697, USA}

\author{Chien-Te Wu}
\author{Abdul N. Malmi-Kakkada}
\author{\\Oriol T. Valls}
\email{otvalls@umn.edu}
\altaffiliation{Also at Minnesota Supercomputer Institute, University of Minnesota,
Minneapolis, Minnesota 55455, USA}
\affiliation{School of Physics and Astronomy, University of Minnesota, 
Minneapolis, Minnesota 55455, USA}

\author{Klaus Halterman }
\email{klaus.halterman@navy.mil}
\affiliation{Michelson Laboratories, Physics
Division, Naval Air Warfare Center, China Lake, California 93555, USA}


\begin{abstract}
We report measurements of the superconducting transition temperature, $T_c$, in CoO/Co/Cu/Co/Nb multilayers as a function of the angle $\alpha$ between the magnetic moments of the Co layers. Our measurements reveal that $T_c(\alpha)$ is a nonmonotonic function, with a minimum near $\alpha={\pi}/{2}$.  Numerical self-consistent solutions of the Bogoliubov--de Gennes equations quantitatively and accurately describe the behavior of $T_c$ as a function of $\alpha$ and layer thicknesses in these superconductor / spin-valve heterostructures. We show that experimental data and theoretical evidence agree in relating $T_c(\alpha)$ to enhanced penetration of the triplet component of the condensate into the Co/Cu/Co spin valve in the maximally noncollinear magnetic configuration.
%

\end{abstract}


\maketitle

\section{INTRODUCTION}
\vspace{-3mm}
Competition between superconducting (S) and ferromagnetic (F) ordering in S/F heterostructures can lead to unusual types of superconductivity emerging 
from the proximity effect at the S/F interfaces \cite{Bulaevskii1977,Buzdin1982,Tokuyasu1988,Ryazanov2001,Kontos2002,Buzdin2005,Bergeret2007,Eschrig2011,Halterman2002}. Penetration of spin-singlet Cooper pairs from the S into the F material can result, when more than one magnetic orientation
is present, in mixing of the spin-triplet and spin-singlet states by the exchange field and generation of a spin-triplet component of the condensate \cite{Buzdin2005,Keizer2006,Bergeret2007,Wang2010,Visani2012,Hubler2012,Halterman2007,Halterman2008}.
The amplitude of this proximity-induced triplet state sensitively depends on the state of magnetization of the F material. In particular, 
the triplet components with nonzero projection of the spin angular momentum of the Cooper pair $\left(S_z=\pm 1\right)$ can only occur when 
there are magnetization noncollinearities. These components of the condensate are immune to pair breaking by the exchange field and, unlike the singlet and the $S_z =0$ triplet components, they can penetrate deep into the F material \cite{Bergeret2007,Halterman2008,Wu2012}. This long-range triplet condensate can be manipulated via changing the relative orientation of the  magnetizations, which creates opportunities for the development of a new class of superconducting spintronic devices. Recent progress in this direction is demonstration of Josephson junctions with noncollinear magnetic barriers, in which the supercurrent is carried by the long-range triplet component of the condensate \cite{Robinson2010,Khaire2010,Anwar2012}. 

Thin-film multilayers of S and F materials are a convenient experimental platform for studies of the proximity-induced triplet condensate \cite{Gu2002,Moraru2006,Gu2010,Leksin2012,Liu2012,Zdravkov2013,Zhu2013,Li2013,Nowak2013}. The advantages of the F/S multilayers include (i) well-established methods of the multilayer deposition, (ii) easy and controllable manipulation of the magnetic state of the F layers via application of external magnetic field, and (iii) convenience of theoretical description of the condensate owing to the translational symmetry in the multilayer plane. Here we present studies of the dependence of $T_c$ in CoO/Co/Cu/Co/Nb multilayers on the in-plane angle $\alpha$ between the magnetic moments of the Co layers.  We compare our experimental results to numerical solutions of the Bogoliubov--de Gennes equations and find that excellent quantitative agreement with the experiment can be achieved when scattering at the multilayer interfaces is taken into account. This solution also reveals that $T_c$ suppression observed for the orthogonal state of the Co magnetic moments originates from enhanced penetration of the long-range triplet condensate into the Co/Cu/Co spin valve in this maximally non-collinear magnetic state. Comparison between the theoretical and experimental $T_c(\alpha)$ allows us to quantify the induced triplet pair amplitude in the spin valve, which reaches values greater than 1\% of the singlet pair amplitude in the Nb layer for the maximally noncollinear ($\alpha={\pi}/{2}$) configuration of the spin valve.

\section{SAMPLE PREPARATION AND CHARACTERIZATION}
\begin{figure}
\centering
\includegraphics[width=0.9\columnwidth]{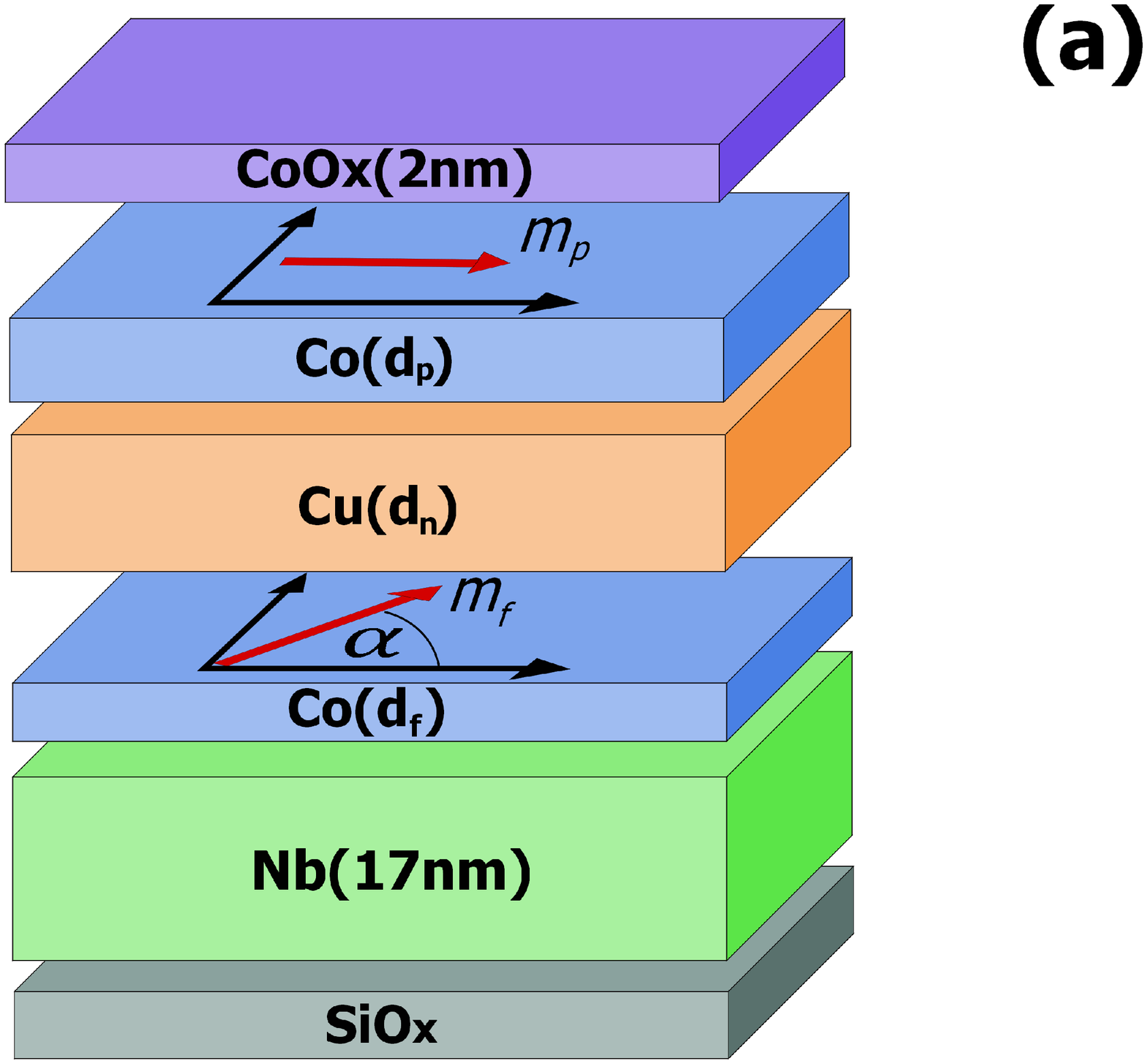}
\includegraphics[width=0.85\columnwidth]{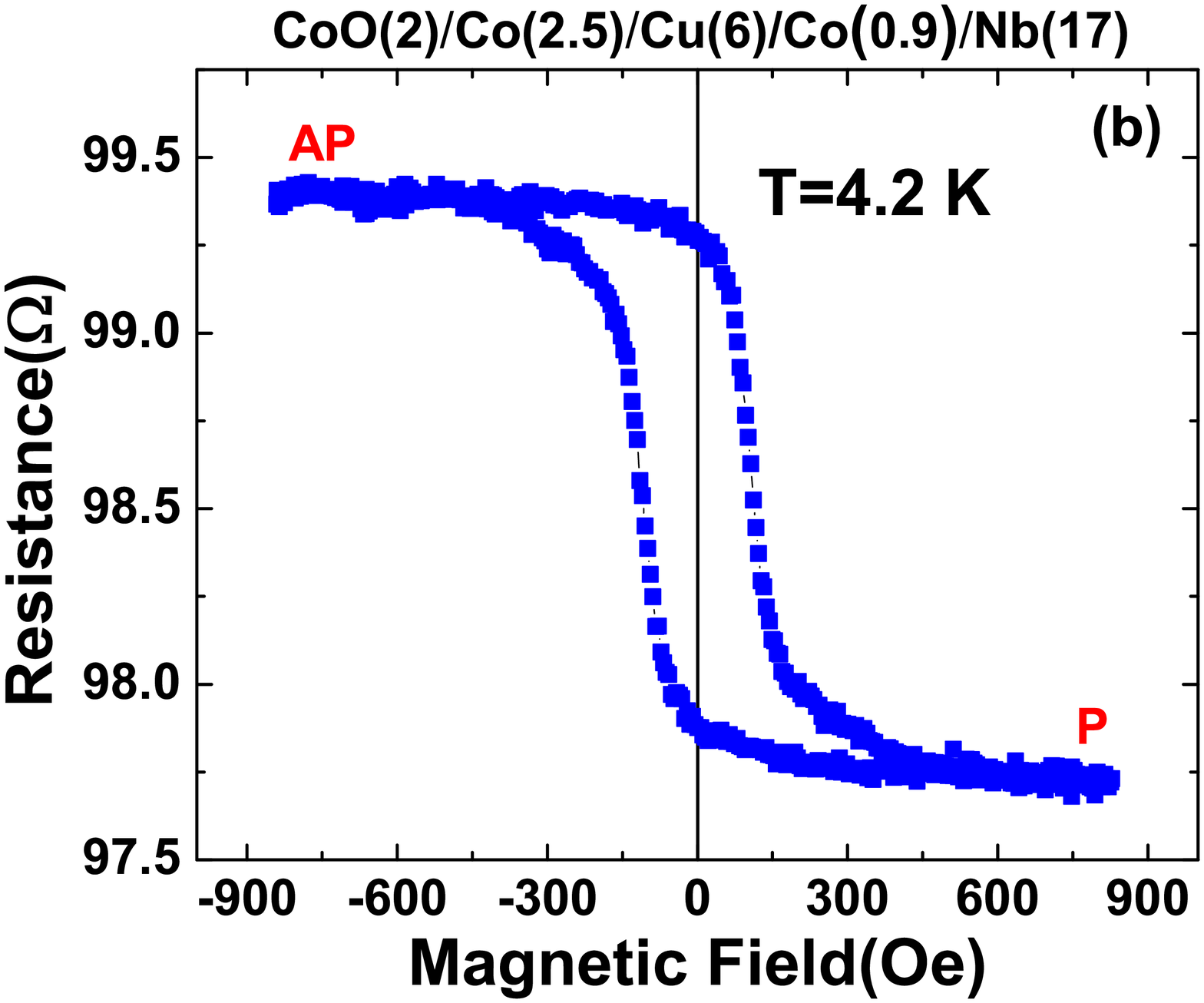}
\vspace{-3mm}
\caption{(Color online) (a) Schematic of the CoO(2 nm)/ Co($d_p$)/ Cu($d_n$)/ Co($d_f$)/ Nb(17 nm) multilayer, where $\alpha$ is the in-plane angle between the magnetic moments of the Co layers. (b) Resistance versus the in-plane magnetic field applied parallel to the pinned layer magnetization at $T$ = 4.2 K (above the superconducting transition temperature).}
\label{RvsH}
\end{figure}

\vspace{-3mm}
The CoO(2 nm)/ Co($d_p$)/ Cu($d_n$)/ Co($d_f$)/ Nb(17 nm)/ (substrate) multilayers, schematically shown in Fig. \ref{RvsH}(a), were prepared by magnetron sputtering in a vacuum system with a base pressure of $8.0\times10^{-9}$ Torr. The deposition was performed onto thermally oxidized Si substrates at room temperature under an Ar pressure of 2 mTorr. The 2 nm thick CoO layer was formed by oxidation of the top part of the Co layer in air for at least 24 hours. The native CoO film is antiferromagnetic at cryogenic temperatures and its purpose is to pin the direction of the top Co layer via the exchange bias phenomenon \cite{Gredig2000}. Three series of multilayers, each series with varying thickness of one of the layers (pinned $d_p$, free $d_f$, and nonmagnetic $d_n$) were deposited in continuous runs with minimal breaks between the samples within the series. This ensured that samples within each of the series were prepared in similar residual gas environments. The three multilayer series reported in this work were designed to elucidate the dependence of the triplet condensate pair amplitude on the spin valve parameters. The description of the series geometries is as follows:

Series 1: CoO(2 nm)/ Co(2.5 nm)/ Cu(6 nm)/ Co($d_f$)/ Nb(17 nm) with $d_{f}$ ranging from 0.5 nm to 1.0 nm  

Series 2: CoO(2 nm)/ Co(2.5 nm)/ Cu($d_n$)/ Co(0.6 nm)/ Nb(17 nm) with $d_n$ ranging from 4 nm to 6.8 nm 

Series 3: CoO(2 nm)/ Co($d_p$)/ Cu(6 nm)/ Co(0.6 nm)/ Nb(17 nm) with $d_p$ ranging from 1.5 nm to 5.5 nm.

The multilayers were patterned into 200 $\mu$m-wide Hall bars using photolithography and liftoff. Four-point resistance measurements of the samples were performed in a continuous flow $^{4}$He cryostat. The magnetization direction of the top Co layer was pinned in the plane of the sample by a strong ($\sim$ 1 T) \cite{Gredig2000} exchange bias field from the antiferromagnetic CoO layer. The exchange bias field direction was set by a 1500 Oe in-plane magnetic field applied to the sample during cooling from the room temperature. As we demonstrate below, the magnetization of the free Co layer can be easily rotated in the plane of the sample by a relatively small ($\sim$ 500 Oe) magnetic field. The role of the nonmagnetic Cu spacer layer is to decouple the magnetic moments of the Co layers, and it is chosen to be thick enough ($d_n>$4 nm) so that both the direct and the RKKY \cite{Parkin1991} exchange interactions between the Co layer are negligibly small. In all magnetoresistance measurements reported here, care is taken to align the applied magnetic field with the plane of the sample so that vortex flow resistance is negligible \cite{Zhu2009}. 

Figure \ref{RvsH}(b) shows the resistance of a CoO(2 nm)/ Co(2.5 nm)/ Cu(6 nm)/ Co(0.9 nm)/ Nb(17 nm) sample as a function of the magnetic field applied along the exchange bias direction measured at $T$ = 4.2 K (above the superconducting transition temperature $T_c$). At $T$ = 4.2 K, all samples show the conventional giant magnetoresistance (GMR) effect originating from the Co/Cu/Co spin valve. Given that there is significant current shunting through the Nb layer, the magnitude of the GMR ($\sim$2\%) is large, demonstrating good quality of both Co/Cu interfaces \cite{Parkin1991}. The GMR curve also demonstrates that external in-plane magnetic field of $\geq$ 500 Oe fully saturates the free layer magnetization along the applied field direction. The lack of an offset in the GMR hysteresis loop from the origin demonstrates that the interlayer exchange coupling between the Co layers is negligible.
\vspace{-5mm}
\section{ANGULAR DEPENDENCE OF $T_c$}
\begin{figure}
\centering
\includegraphics[width=0.85\columnwidth]{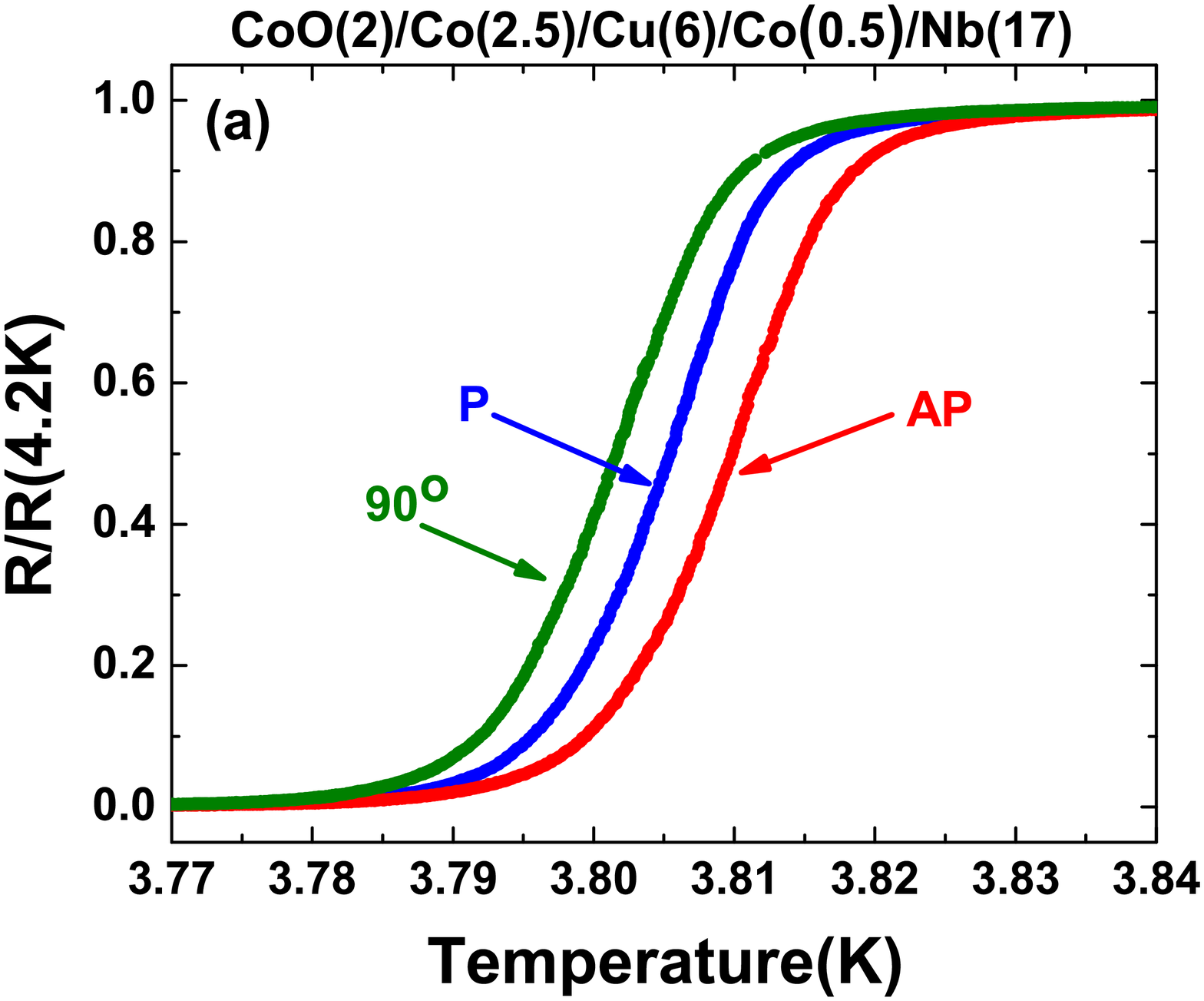}
\includegraphics[width=0.85\columnwidth]{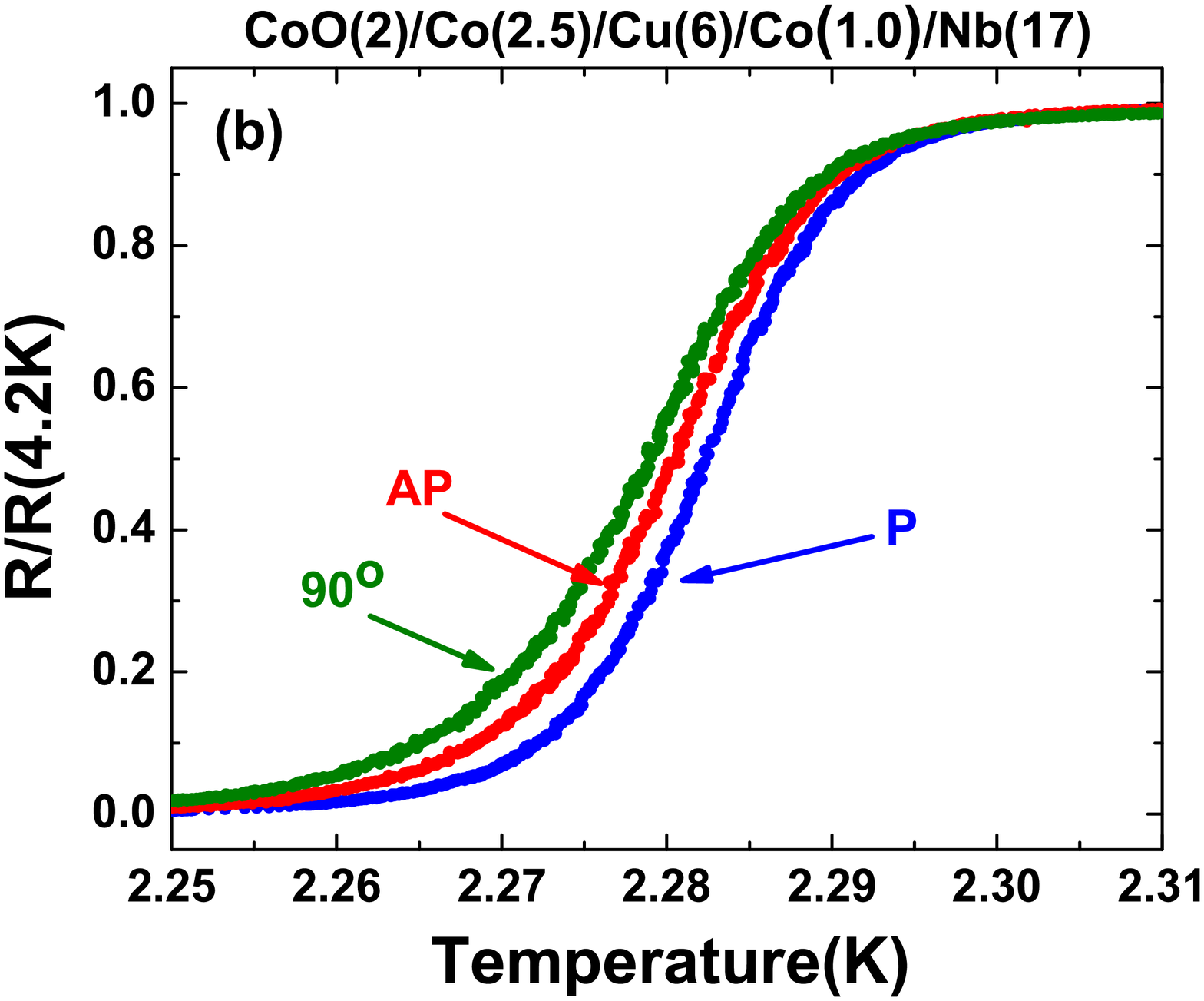}
\vspace{-2mm}
\caption{(Color online) Resistance versus temperature for parallel (P, $\alpha = 0$), antiparallel (AP, $\alpha = \pi$), and perpendicular ($90^\circ$, $\alpha = \frac{\pi}{2}$) orientations of magnetic moments of the Co layers for multilayer samples with (a) $d_f=$ 0.5 nm and (b) $d_f=$ 1.0 nm. The resistance is divided by its normal state value measured at $T =$ 4.2 K.}
\label{RvsT}
\end{figure}

\vspace{-3mm}
We next make measurements of the multilayer superconducting transition temperature $T_c$ as a function of the angle $\alpha$ between magnetic moments of the pinned and free layers. We define $T_c$ as the temperature at which the sample resistance becomes equal to half of its normal state value. For these measurements, we use an 800 Oe in-plane magnetic field to set the direction of magnetization of the free layer. As discussed in the previous section, this field completely saturates the magnetization of the free layer in the direction of the field. Furthermore, this field is much smaller than the exchange bias field acting on the pinned layer and thus we assume that the pinned layer magnetization remains in the direction of the cooling field for all our measurements. Figure \ref{RvsT} shows resistance versus temperature measured in the parallel (P, $\alpha=0$), antiparallel (AP, $\alpha=\pi$), and perpendicular (90$^\circ$, $\alpha={\pi}/{2}$) configurations of the two Co layers for the samples with 0.5 nm and 1.0 nm thick Co free layers, and 2.5 nm thick Co pinned layer. In this measurement, the angle between the magnetic moments is pinned by the in-plane external field while the temperature is swept across the superconducting transition. To ensure that the sample remains in thermal equilibrium with the bath, the temperature for each measurement is swept at a sufficiently slow rate of 2 mK per minute. For both values of the free layer thickness, we find that the perpendicular configuration of the spin valve ($\alpha={\pi}/{2}$) gives the lowest transition temperature $T_c$. We find this to be  universally true for all samples studied in this work: $T_c({\pi}/{2}) < T_c(0)$ and $T_c({\pi}/{2}) < T_c(\pi)$. In contrast, the relation between $T_c$ in the P and AP configurations depends on the thickness of the free layer. Figure \ref{RvsT} shows that $T_c(0) < T_c(\pi)$ for $d_f=$ 0.5 nm, while $T_c(\pi) < T_c(0)$ for $d_f=$ 1.0 nm. Similar trends in the angular and thickness dependence of $T_c$ were recently observed in Pb/ Fe/ Cu/ Fe/ CoO multilayers \cite{Leksin2012}. 

\begin{figure}
\centering
\includegraphics[width=0.85\columnwidth]{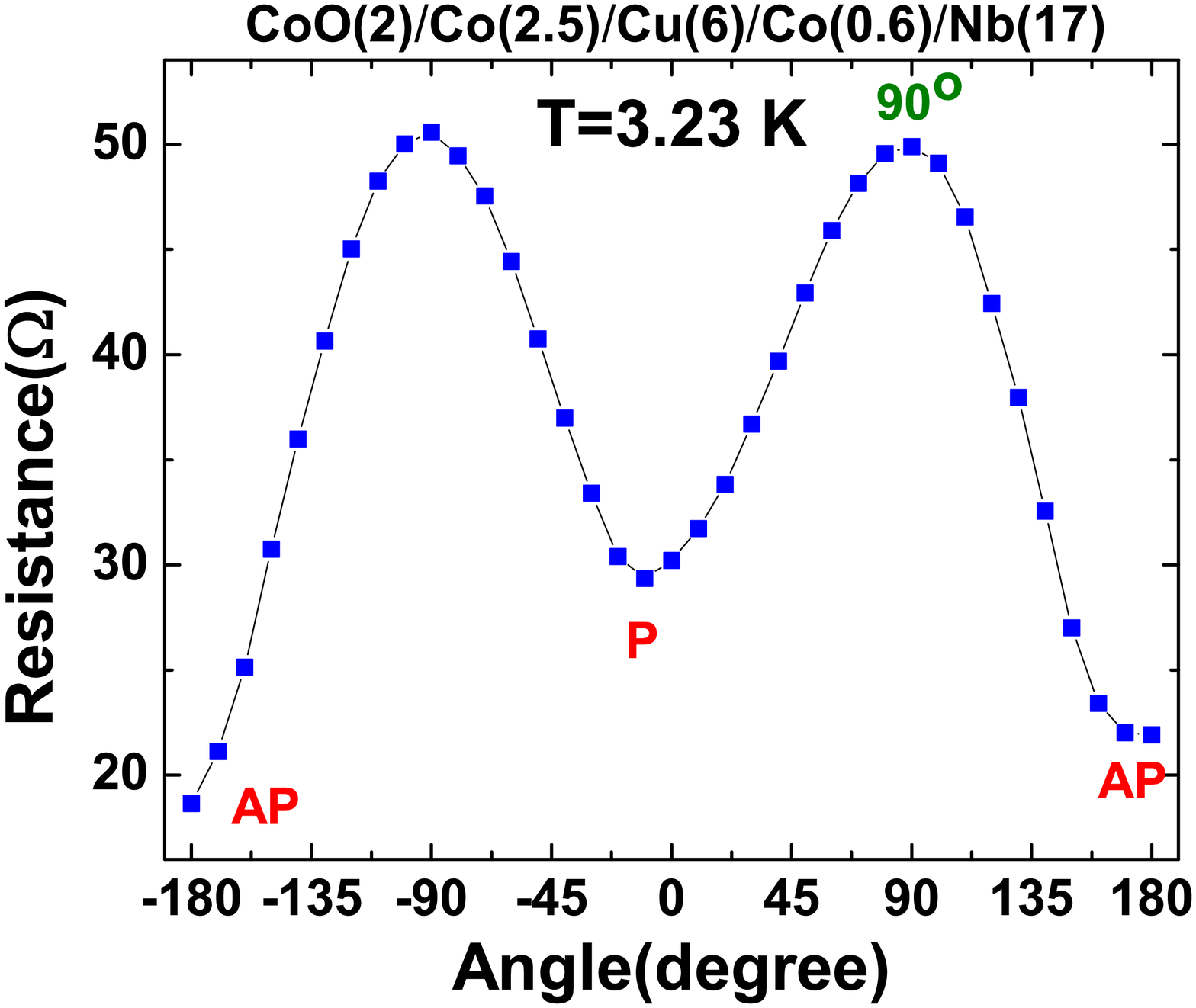}
\vspace{-2mm}
\caption{(Color online) Resistance of a CoO(2 nm)/ Co(2.5 nm)/ Cu(6 nm)/ Co(0.6 nm)/ Nb(17 nm) structure versus magnetic field angle, $\alpha$ , measured at $T =$ 2.92 K in the middle of the superconducting transition, at a field of 800 Oe.}
\label{RvsA}
\end{figure}

To understand the angular dependence of $T_c$ in greater detail, we fix the temperature in the middle of the superconducting transition and measure the sample resistance $R$ as a function of in-plane angle $\alpha$ between the magnetic moments of the pinned and free layers. This measurement is made by applying an 800 Oe saturating magnetic field and rotating it through 360$^\circ$ in the plane of the sample. Figure \ref{RvsA} shows $R(\alpha)$ measured at $T=$ 2.92 K (the middle of the superconducting transition) for a CoO(2 nm)/ Co(2.5 nm)/ Cu(6 nm)/ Co(0.6 nm)/ Nb(17 nm) sample. Because resistance is a steep function of temperature in the middle of the superconducting transition, we take great care to stabilize the temperature to within $\pm 0.1$ mK during these measurements in order to reduce the level of thermal noise in the $R(\alpha)$ data.

\begin{figure}[H]
\centering
\includegraphics[width=0.85\columnwidth]{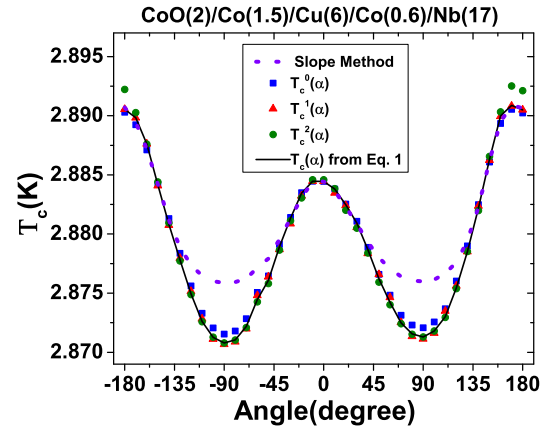}
\vspace{-2mm}
\caption{(Color online) $T_c(\alpha)$ for a CoO(2 nm)/ Co(1.5 nm)/ Cu(6 nm)/ Co(0.6 nm)/ Nb(17 nm) multilayer calculated from the $R(\alpha)$ data by different methods described in the text.}
\label{TcvsA}
\end{figure}

Measurements of $R(\alpha)$ are much faster than those of $R(T)$ because reliable $R(T)$ data require sweeping temperature at slow rates. Thus we employ the $R(\alpha)$ data in order to evaluate $T_c(\alpha)$ instead of direct measurements of $T_c(\alpha)$ at multiple values of $\alpha$, such as those shown in Fig. \ref{RvsT}. We therefore need a reliable method of extracting $T_c(\alpha)$ from the $R(\alpha)$ data. The simplest method for such extraction is to use the slope of the $R(T)$ curve at $T_c$ for $\alpha=$ 0 and to calculate $T_c(\alpha)$ as $T_c(\alpha)=T_c(0)-({dT}/{dR})\left[R(\alpha)-R(0)\right]$, 
where $R(\alpha)$ is the experimentally measured angular dependence of resistance at $T=T_c(0)$. This simple method assumes approximately linear variation of resistance with temperature near $T_c$ and already gives qualitatively satisfactory results.  However, the maximum uncertainty in the resulting $T_c(\alpha)$ can be as large as 5 mK. The purple dotted curve in Fig. (\ref{TcvsA}) shows $T_c(\alpha)$ calculated by this method for a CoO(2 nm)/ Co(2.5 nm)/ Cu(6 nm)/ Co(0.6 nm)/ Nb(17 nm) multilayer. 

In order to take into account deviations of $R(T)$ from a linear function and thereby improve the procedure for extracting $T_c(\alpha)$ from the $R(\alpha)$ data, we calculate $T_c(\alpha)$ based on the experimentally measured $R(T,0)$ and $R(T^*,\alpha)$ curves, where $T^*\approx T_c(0)$ is the temperature at which the angular dependence of resistance is measured. In this method, we assume that the shape of the $R(T)$ curve is the same for all values of $\alpha$, and that the curves at different $\alpha$ can be obtained by simply translating the experimentally measured $R(T,0)$ curve along the temperature axis by $\Delta T_c(\alpha)=T_c(\alpha)-T_c(0)$. With this assumption, $T_c(\alpha) = T_c(0)+\Delta T_c(\alpha)$ can be found by numerically solving the implicit equation $R(T^*,\alpha)=R(T^*-\Delta T_c(\alpha),0)$ for $\Delta T_c(\alpha)$. The blue squares in Fig. (\ref{TcvsA}) show $T_c(\alpha)$ evaluated by this method using the transition curve from the P ($\alpha=0$) state. The red triangles and green dots represent the same method used with the other two measured curves.  We find this method of evaluating $T_c(\alpha)$ to be quite reliable for our samples with an estimated error of $\sim$ 1 mK.

\begin{figure*}[htb]
\centering
\includegraphics[width=0.73244\columnwidth]{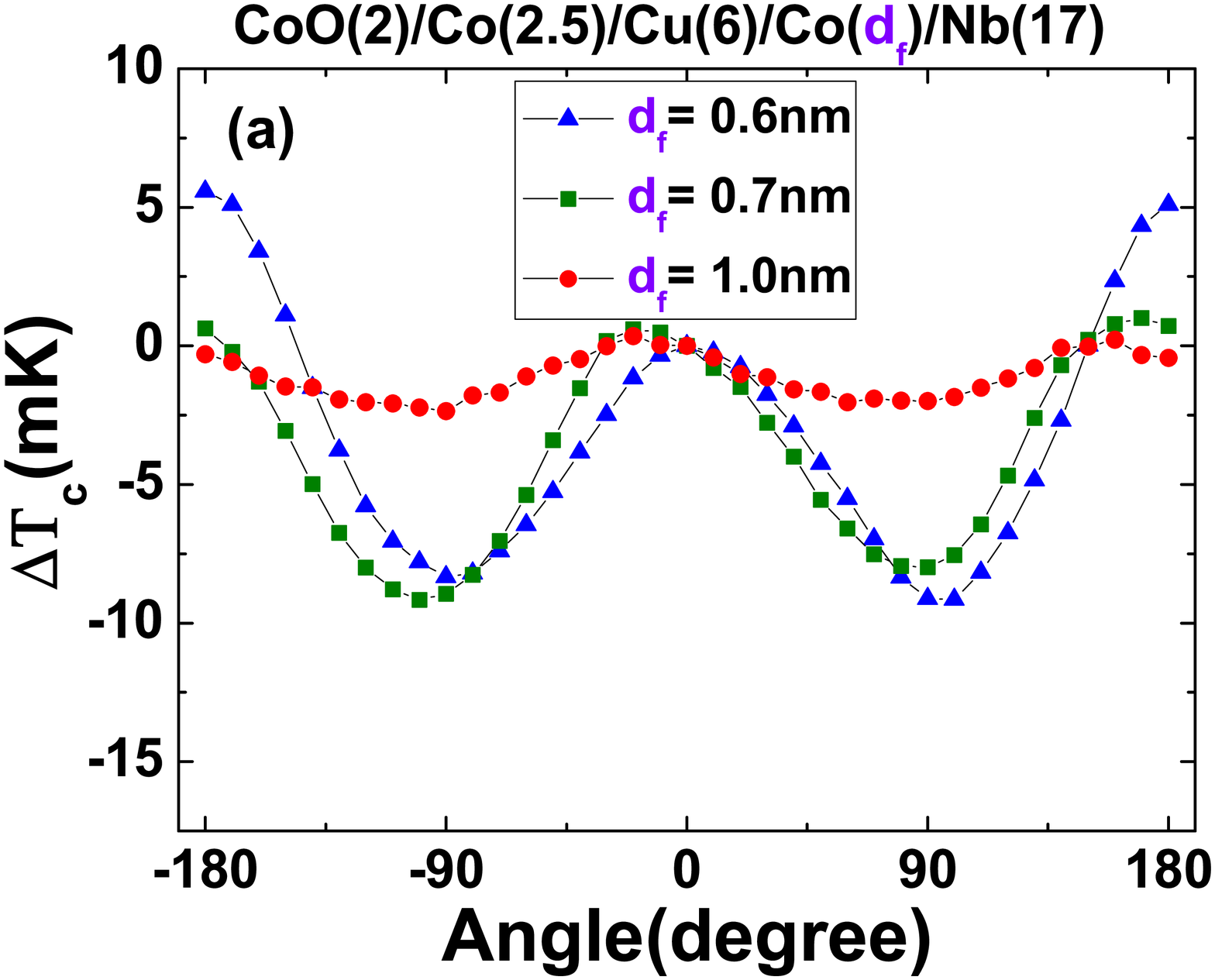}
\includegraphics[width=0.63378\columnwidth]{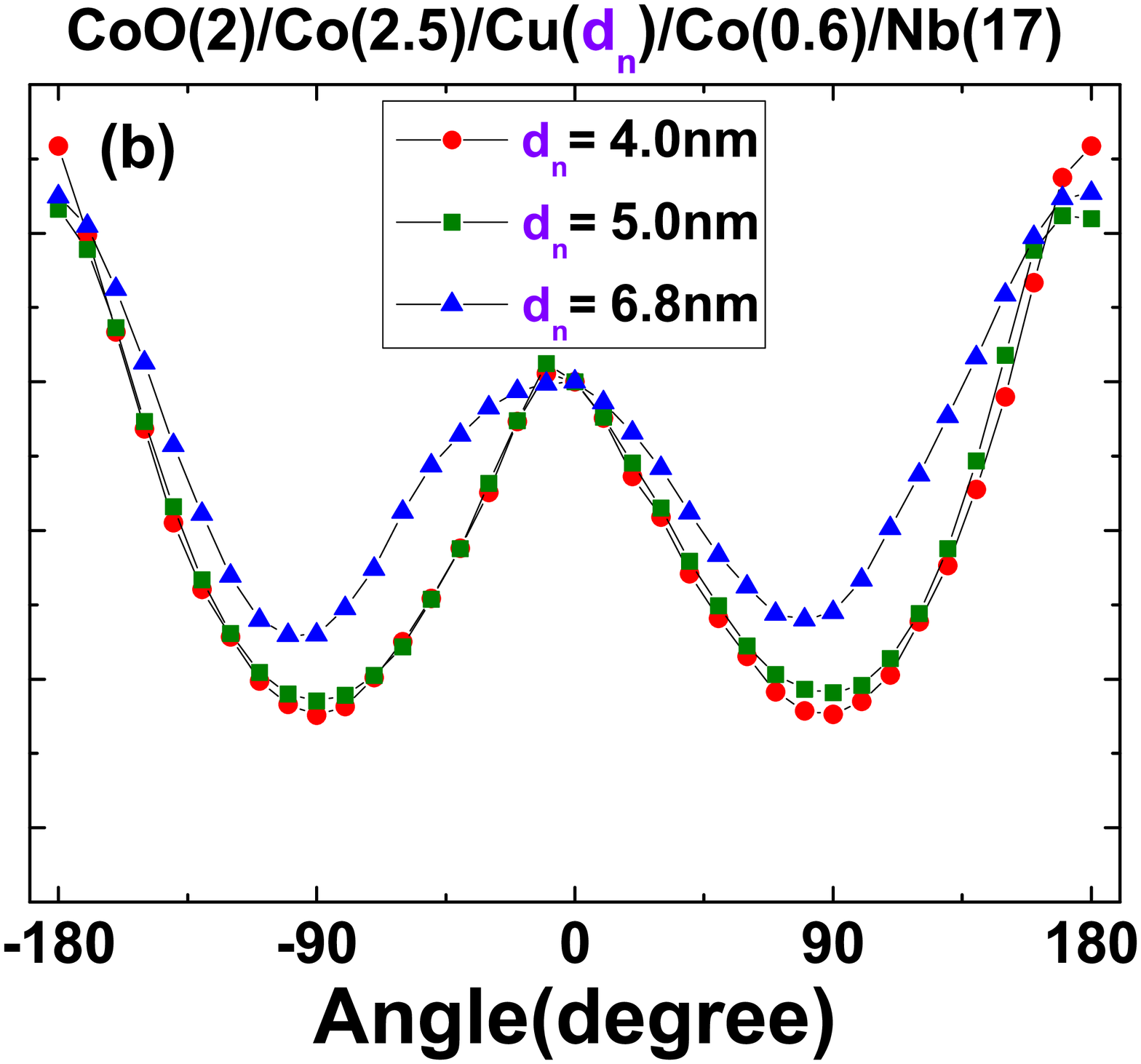}
\includegraphics[width=0.63378\columnwidth]{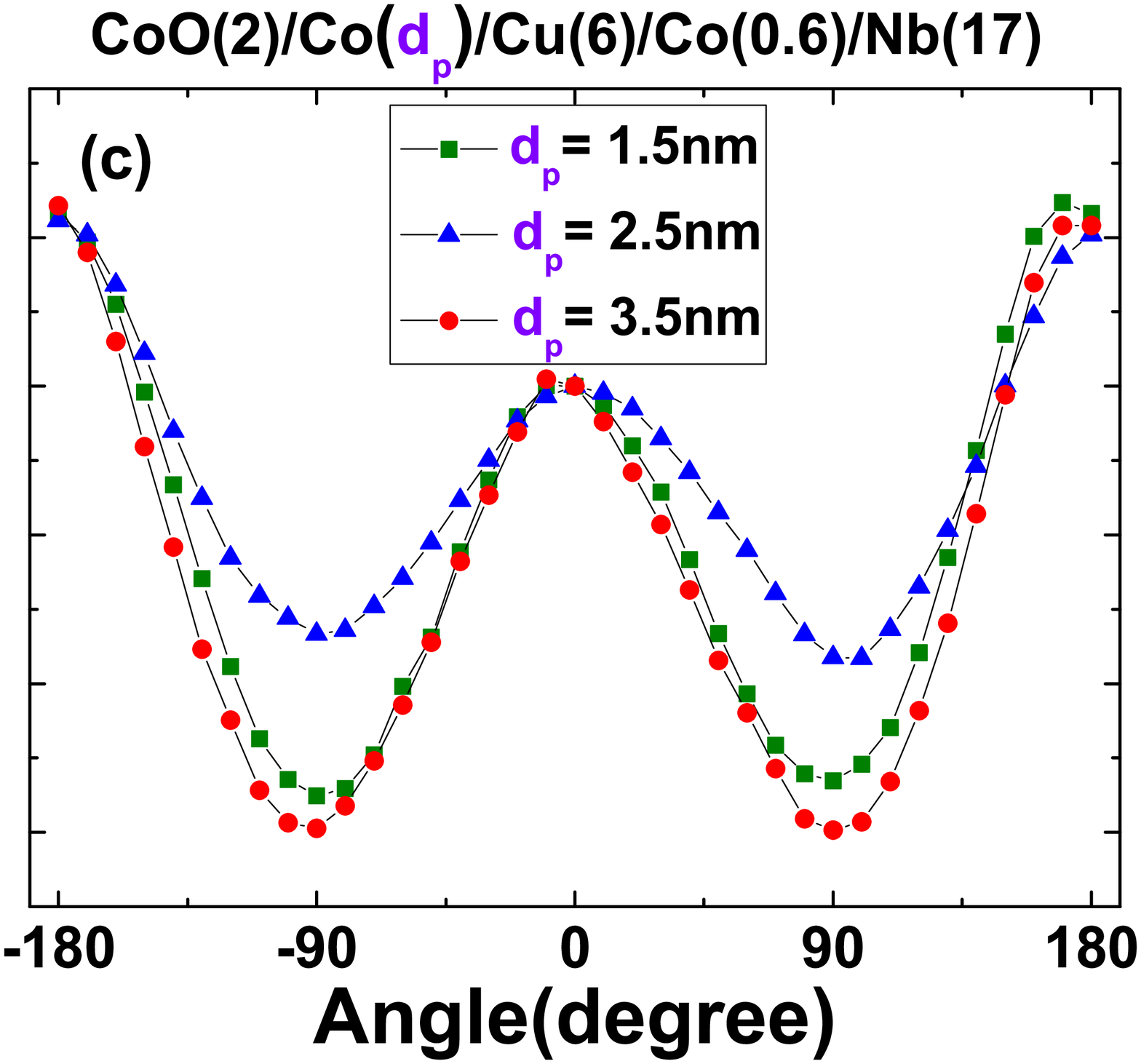}
\vspace{-2mm}
\caption{(Color online) $T_c(\alpha)$ for representative samples from the three series of samples studied in this paper. (a) From the $d_{f}$ series, (b) from the $d_n$ series, (c) from the $d_{p}$ series.}
\label{DTcvsA}
\end{figure*}

An even more refined method of evaluating $T_c(\alpha)$ takes into account that the shape of the $R(T)$ curve (not only its position along the temperature axis) may depend on $\alpha$. Here we first calculate $\Delta T_c(\alpha)$ based on the experimentally measured $R\left(T,{\pi}/{2}\right)$ and $R(T,\pi)$ curves using the method described above: we calculate $T_c(\alpha)$ by numerically solving the implicit equations $R(T^*,\alpha)=R\left(T^*-\Delta T_c(\alpha),n{\pi}/{2}\right)$ for $\Delta T_c(\alpha)$, where $\Delta T_c(\alpha)=T_c(\alpha)-T_c\left(n{\pi}/{2}\right)$, $n=1,2$. These $T_c(\alpha)$ values calculated for $n=1,2$ are shown in Fig. (\ref{TcvsA}) as green circles and red triangles, respectively. Figure (\ref{TcvsA}) clearly illustrates that all three functions $T_c(\alpha)$ calculated by numerically solving the implicit equations written above for $n=0,1,2$ are very similar to each other. The average of these three functions $T_c^n(\alpha)$, which we now explicitly label by the index $n=0,1,2$, would give a reasonable result for $T_c(\alpha)$. However, a better estimate is given by the following equation:
\begin{equation}
T_c(\alpha) = \sum_{n=0}^2 T_c^n(\alpha) w_n(\alpha)
\label{efun}
\end{equation}

\begin{figure*}[htb]
\centering
\includegraphics[width=0.73244\columnwidth]{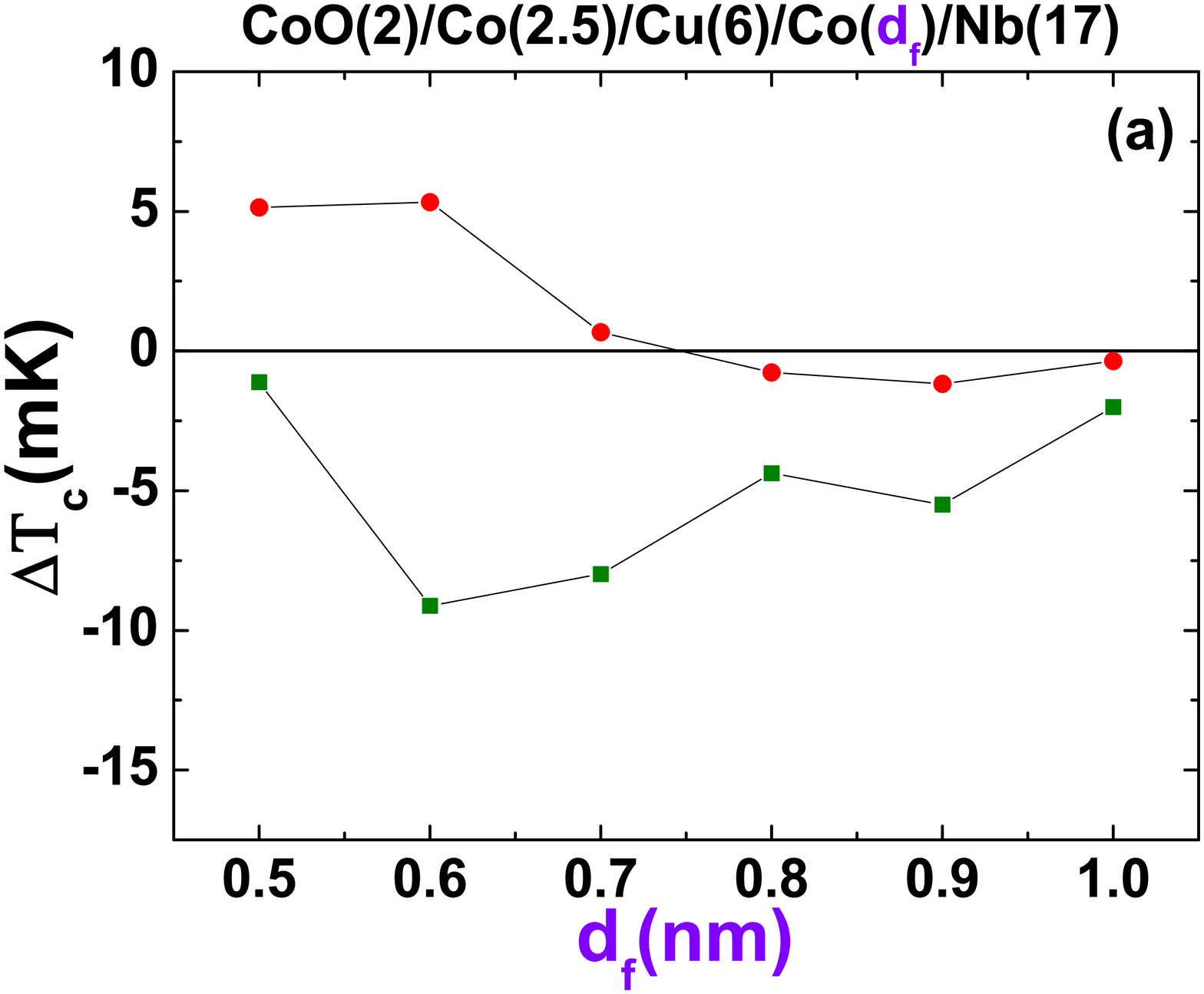}
\includegraphics[width=0.63378\columnwidth]{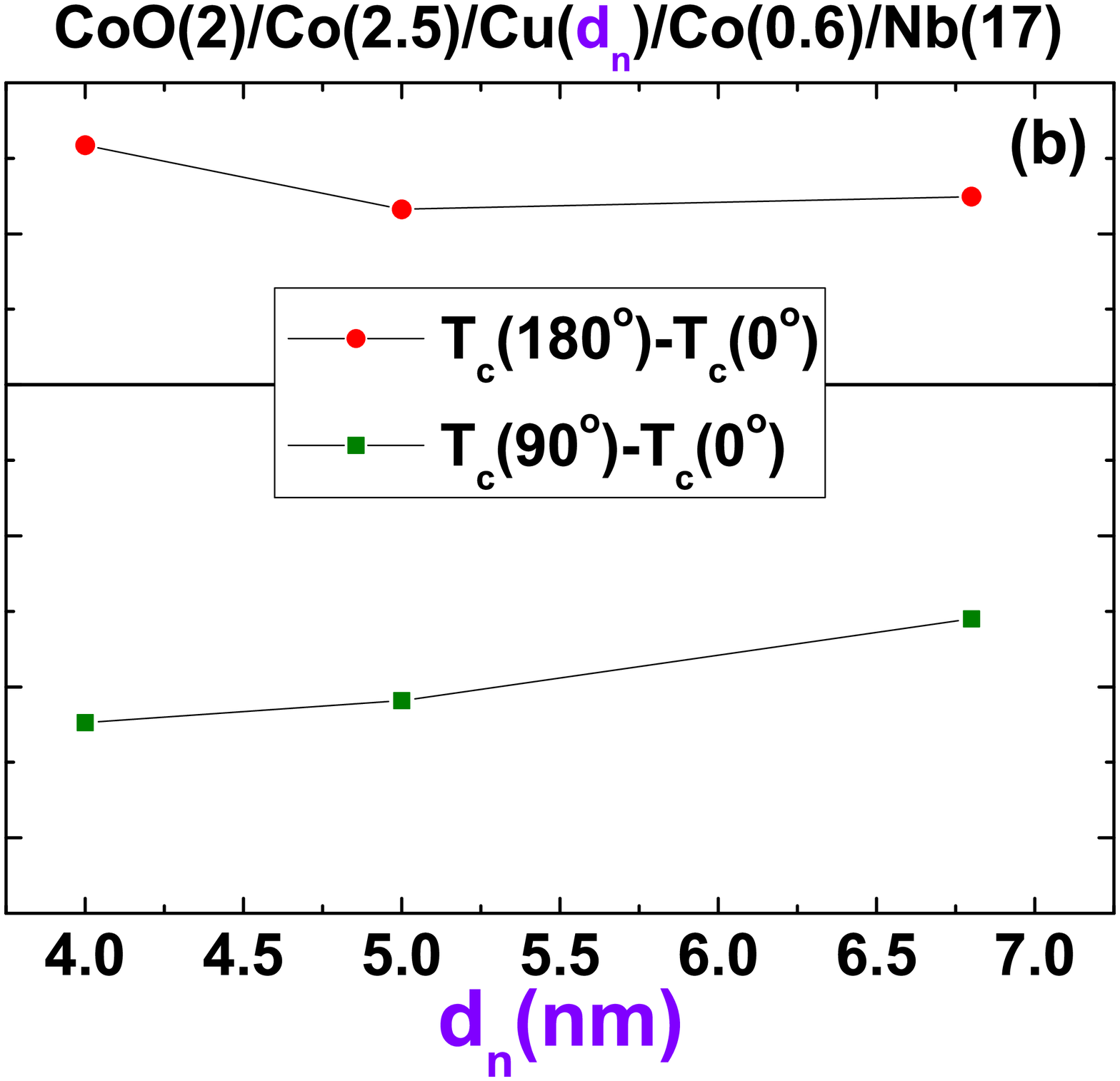}
\includegraphics[width=0.63378\columnwidth]{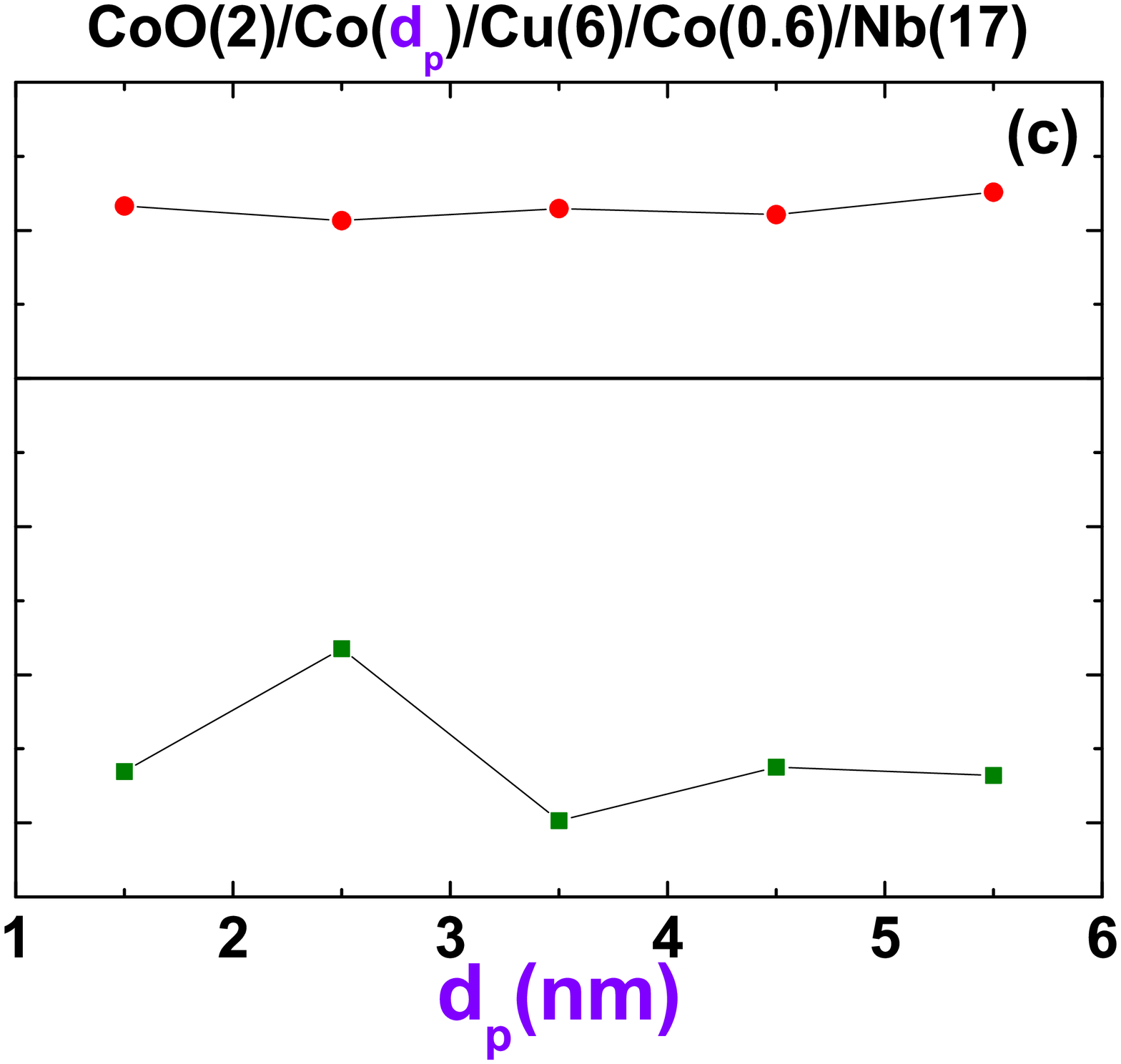}
\vspace{-2mm}
\caption{(Color online) Dependence of $T_c(\pi)-T_c(0)$ (red circles) and $T_c(\frac{\pi}{2})-T_c(0)$ (green squares) on the free Co layer thickness $d_f$ (a), nonmagnetic spacer thickness $d_n$ (b), and pinned layer thickness $d_p$ (c).}
\label{DTc_thicknesses}
\end{figure*}

where $w_n(\alpha)$ are extrapolation functions with maxima at $\alpha=\pm n{\pi}/{2}$. The extrapolation functions also satisfy the normalization condition $\sum_{n=0}^2 w_n(\alpha)=1$ on the interval of $\alpha$ from $-\pi$ to $\pi$. We make the following choice of the extrapolation functions: $w_0(\alpha)=\cos^2(\alpha) \Theta\left({\pi}/{2}-\left|\alpha\right|\right)$, $w_1(\alpha)=\sin^2(\alpha)$ and $w_2(\alpha)=\cos^2(\alpha) \Theta\left(\left|\alpha\right|-{\pi}/{2}\right)$, where $\Theta(x)$ is the Heaviside step function. The advantage of Eq. (\ref{efun}) over the simple average is that at $\alpha = 0, {\pi}/{2}, \pi$, the expression for $T_c(\alpha)$ reduces to the exact value of $T_c$ directly measured at these angles in the $R(T)$ measurements. The black solid line in Fig. (\ref{TcvsA}) shows $T_c(\alpha)$ evaluated by this method. We use this method for calculating $T_c(\alpha)$ from the experimental data throughout the rest of the paper.

Figure \ref{DTcvsA} shows a representative angular variation of $T_c$, $\Delta T_c(\alpha)=T_c(\alpha)-T_c(0)$, for the three series of samples employed in our study. We find that for all samples employed in our experiment, $T_c(\alpha)$ is a nonmonotonic function in the interval of $\alpha$ from $-\pi$ to $\pi$ with a minimum near perpendicular orientation of the free and pinned layers ($\alpha={\pi}/{2}$). As we demonstrate in the analysis section of this paper, this minimum in $T_c$ arises from the enhanced long-range triplet pair amplitude in the maximally noncollinear configuration of the spin valve. We also note that our previous studies of the angular dependence of $T_c$ in NiFe/Nb/NiFe trilayers \cite{Zhu2010} found monotonic dependence of $T_c$ on $\alpha$ in the 0 to $\pi$ range, which serves as indication of a much weaker triplet pair amplitude induced in the system with two ferromagnetic layer separated by a superconductor.

Figure \ref{DTc_thicknesses} summarizes the dependence of $T_c(\alpha)$ on the thickness of the Co/Cu/Co spin valve layers. Figure \ref{DTc_thicknesses}(a) shows the difference of $T_c$ in the P and AP states as a function of the free layer thickness $d_f$. The data demonstrate that $T_c(\pi)-T_c(0)$ oscillates and changes sign as a function of $d_f$, which is a consequence of interference of the pair wave function in the free layer. Figures \ref{DTc_thicknesses}(b) and \ref{DTc_thicknesses}(c) show the dependence of $T_c(\pi)-T_c(0)$ on the nonmagnetic spacer thickness $d_n$ and the pinned layer thickness $d_p$. This dependence is weak, which implies that (i) the pair amplitude decays slowly in the Cu spacer layer and (ii) the pair amplitude decays to nearly zero over the pinned layer thickness greater than 1.5 nm (the thinnest pinned layer employed in our studies). 
The behavior of $T_c$ will be discussed in general later in this work. 

Figure \ref{DTc_thicknesses} also illustrates the thickness dependence of $T_c$ in the maximally noncollinear geometry of $\alpha = {\pi}/{2}$. The green squares in Fig. \ref{DTc_thicknesses} show the dependence of $T_c({\pi}/{2})-T_c(0)$ on the spin valve layer thicknesses. This figure clearly shows that $T_c({\pi}/{2})$ is always lower than $T_c(0)$ and $T_c(\pi)$. Figure \ref{DTc_thicknesses}(c) illustrates that $T_c({\pi}/{2})$ shows variation with the pinned layer thickness for $d_p$ as large as 3.5 nm. This serves as evidence of the long-range ($>3.5$ nm ) penetration of the triplet component of the condensate into the pinned ferromagnetic layer. 

\section{THEORETICAL METHODS}\label{methods}
\vspace{-3mm}
The theoretical method we adopted is thoroughly discussed in Refs.~\cite{Halterman2002, Barsic2007, Halterman2008}; therefore, we only present here the essential parts that are necessary for our discussion. In particular, the theoretical method we used to find $T_c$ can be found in Refs.~\cite{Zhu2010,Barsic2007}. We modeled the Nb/Co/Cu/Co heterostructures as ${\rm S/F_f/N/F_p}$ layered systems, where $\rm S$ represents the superconducting layer, $\rm F_f$ and $\rm F_p$ are the inner (free) and outer (pinned) magnets, and $\rm N$ denotes the normal metallic intermediate layer. The layers are assumed to be infinite in the $x$-$z$ plane with a total thickness $d$ in the $y$ direction, which is perpendicular to the interfaces between layers. In accordance with the experiment, $\rm F_p$ has width $d_{p}$, and fixed direction of magnetization. The normal layer with width $d_n$ is sandwiched between this pinned layer and a magnetic layer $\rm F_f$ of width $d_{f}$ with experimentally controlled magnetization direction. The superconducting layer of thickness $d_S$ is in contact with the free layer. The in-plane magnetizations in the $\rm F$ layers are modeled by effective Stoner-type exchange fields ${\bf h}(y)$ which vanish in the nonferromagnetic layers. To accurately describe the physical properties of our systems with sizes in the nanometer scale and moderate exchange fields, where semiclassical approximations are inappropriate, we numerically solve the microscopic Bogoliubov--de Gennes (BdG) equations in a fully self-consistent manner. The geometry of our system allows one to express the BdG equations in a quasi-one-dimensional form (natural units $\hbar=k_B=1$ are assumed),
\begin{widetext}
\begin{align}
\begin{pmatrix} 
{\cal H}_0 -h_z(y)&-h_x(y)&0&\Delta(y) \\
-h_x(y)&{\cal H}_0 +h_z(y)&\Delta(y)&0 \\
0&\Delta(y)&-({\cal H}_0 -h_z(y))&-h_x(y) \\
\Delta(y)&0&-h_x(y)&-({\cal H}_0+h_z(y)) \\
\end{pmatrix} 
\begin{pmatrix} 
u_{n\uparrow}(y)\\u_{n\downarrow}(y)\\v_{n\uparrow}(y)\\v_{n\downarrow}(y)
\end{pmatrix} 
=\epsilon_n
\begin{pmatrix}
u_{n\uparrow}(y)\\u_{n\downarrow}(y)\\v_{n\uparrow}(y)\\v_{n\downarrow}(y)
\end{pmatrix}\label{bogo2},
\end{align}
\end{widetext}
where $h_i(y)$ ($i=x,z$) are components of the exchange fields ${\bf h}(y)$. In Eq.~(\ref{bogo2}), the single-particle Hamiltonian ${\cal H}_0=
-{1}/{(2m)}{d^2}/{dy^2}-E_F+U(y)$ contains the Fermi energy, $E_F$, and an effective interfacial scattering potential described by delta functions of strength $H_j$ ($j$ denotes the different interfaces); namely,
\begin{align}
U(y)= &H_{1}\delta(y-d_S)+H_{2}\delta(y-d_S-d_{f}) \nonumber \\
&+H_{3}\delta(y-d_S-d_{f}-d_n), 
\label{scat}
\end{align}
where $H_j={k_FH_{Bj}}/{m}$ is written in terms of the dimensionless scattering strength $H_{Bj}$. We assume $h_x(y)=h_0\sin(-\alpha/2)$ and $h_z(y)=h_0\cos(-\alpha/2)$  in $\rm F_f$, where $h_0$ is the magnitude of exchange field. In $\rm F_p$, we have $h_x(y)=h_0\sin(\alpha/2)$ and \cite{Wu2012} $h_z(y)=h_0\cos(\alpha/2)$. The functions $u_{n\sigma}$ and $v_{n\sigma}$ ($\sigma=\uparrow$, $\downarrow$) in Eq.~(\ref{bogo2}) represent quasiparticle and quasihole wave functions. By applying the generalized Bogoliubov transformations (see Ref.~\cite{Halterman2008}), the self-consistent singlet pair potential $\Delta(y)$ can be expressed in terms of quasiparticle and quasihole wave functions; that is,
\begin{equation}   
\label{del} 
\Delta(y) = \frac{g(y)}{2}{\sum_n}^\prime
\bigl[u_{n\uparrow}(y)v_{n\downarrow}(y)+
u_{n\downarrow}(y)v_{n\uparrow}(y)\bigr]\tanh\Bigl(\frac{\epsilon_n}{2T}\Bigr), \,
\end{equation} 
where the primed sum means summing over all  eigenstates with energies $\epsilon_n$ that lie within a characteristic Debye energy $\omega_D$, and $g(y)$ is the superconducting coupling strength, taken to be constant in the $\rm S$ region and zero elsewhere. We have assumed that the quantization axis lies along the $z$ direction, but one can easily obtain the spin-dependent quasiparticle amplitudes with respect to a different spin quantization axis rotated by an angle $\theta$ in the $x$-$z$ plane via the rotation matrix \cite{Halterman2008}:
\begin{equation}   
\label{rotation}
\hat{U_0}(\theta)=\cos(\theta/2)\hat{\mathrm{I}}\otimes\hat{\mathrm{I}}-i\sin(\theta/2)\rho_z\otimes\sigma_z,
\end{equation}
where $\bm \rho$ and $\bm \sigma$ are vectors of Pauli matrices in particle-hole and spin space, respectively. 

\begin{figure*}[htb]
\centering
\includegraphics[width=0.63378\columnwidth]{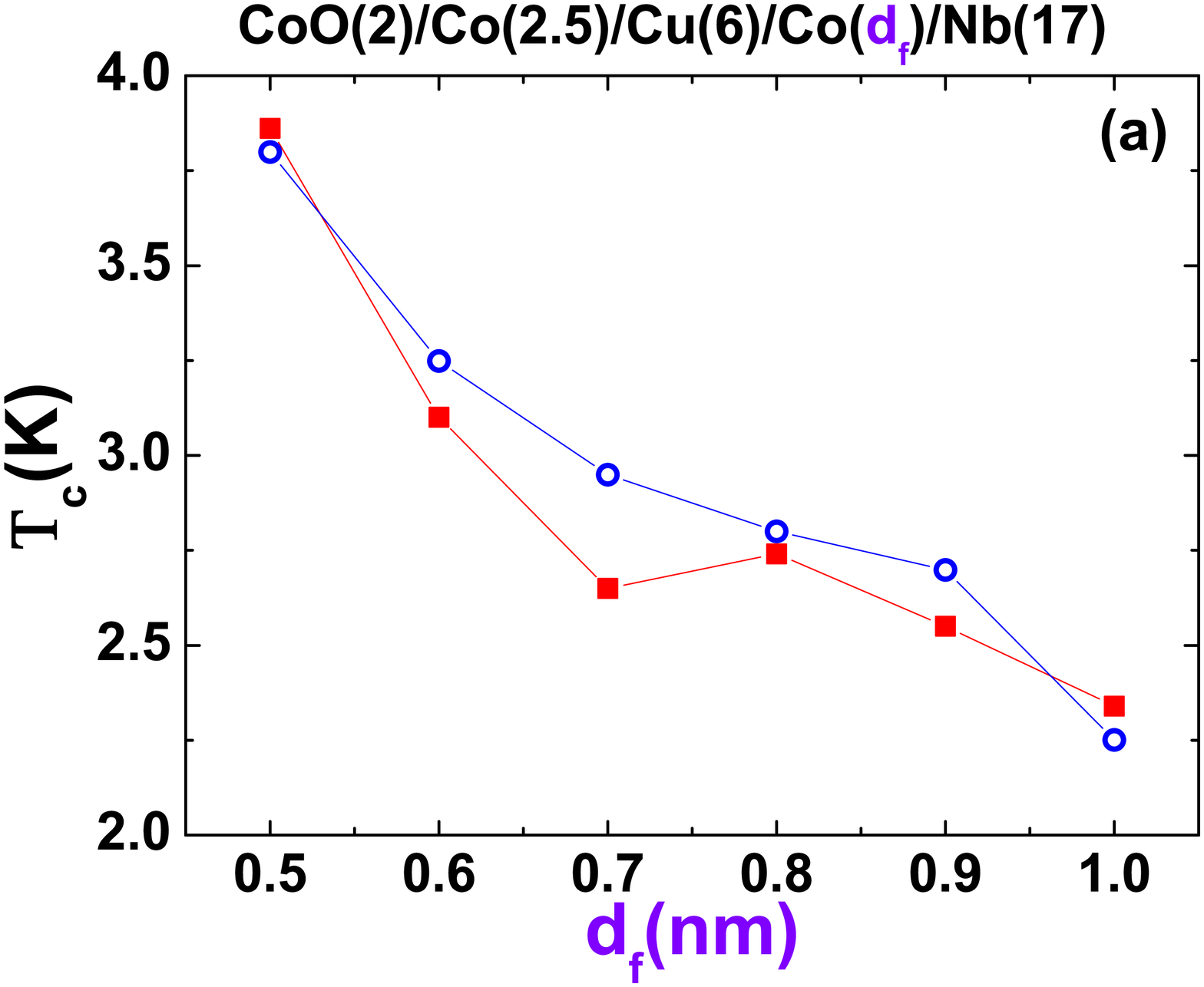} 
\includegraphics[width=0.63378\columnwidth]{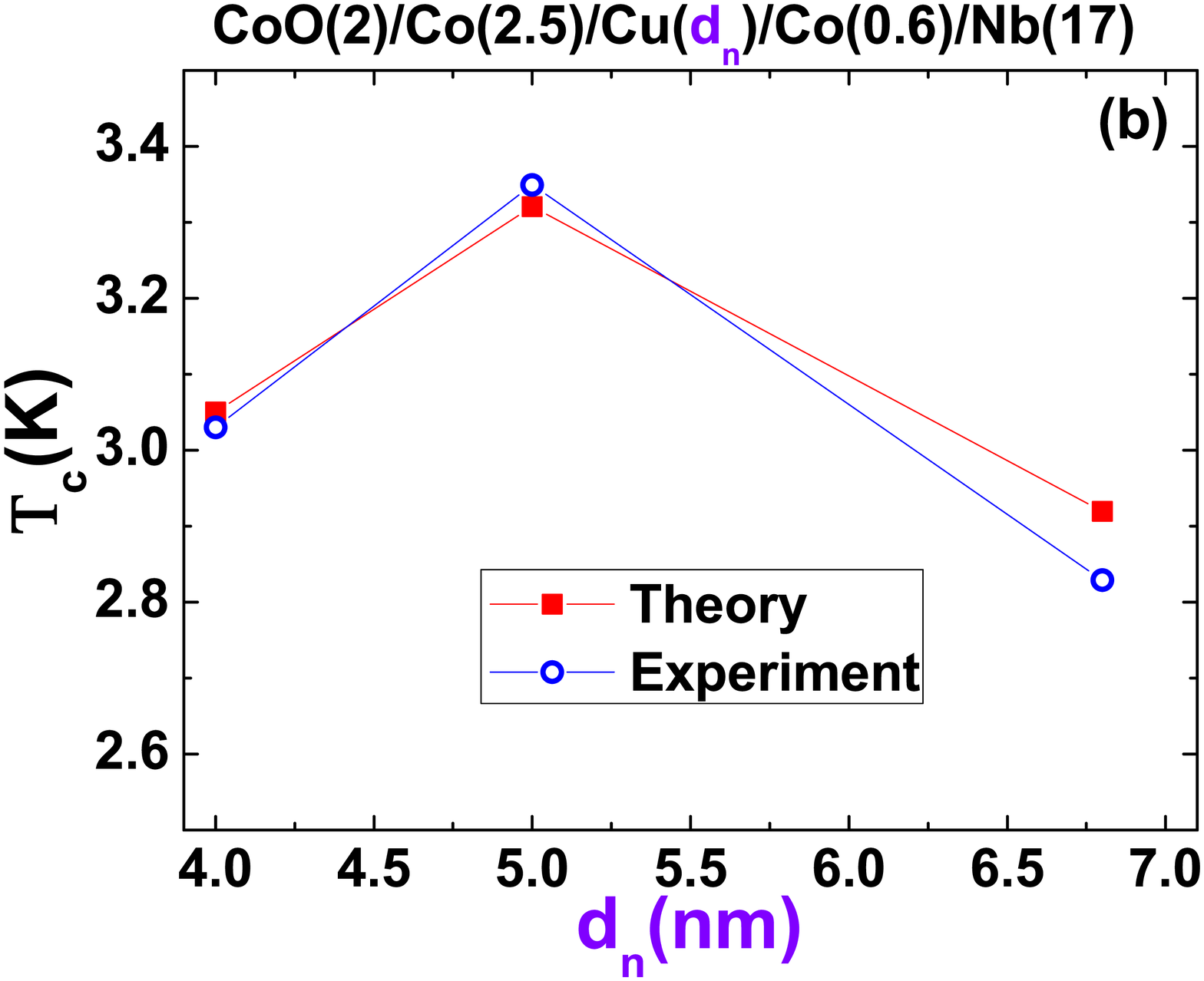}
\includegraphics[width=0.63378\columnwidth]{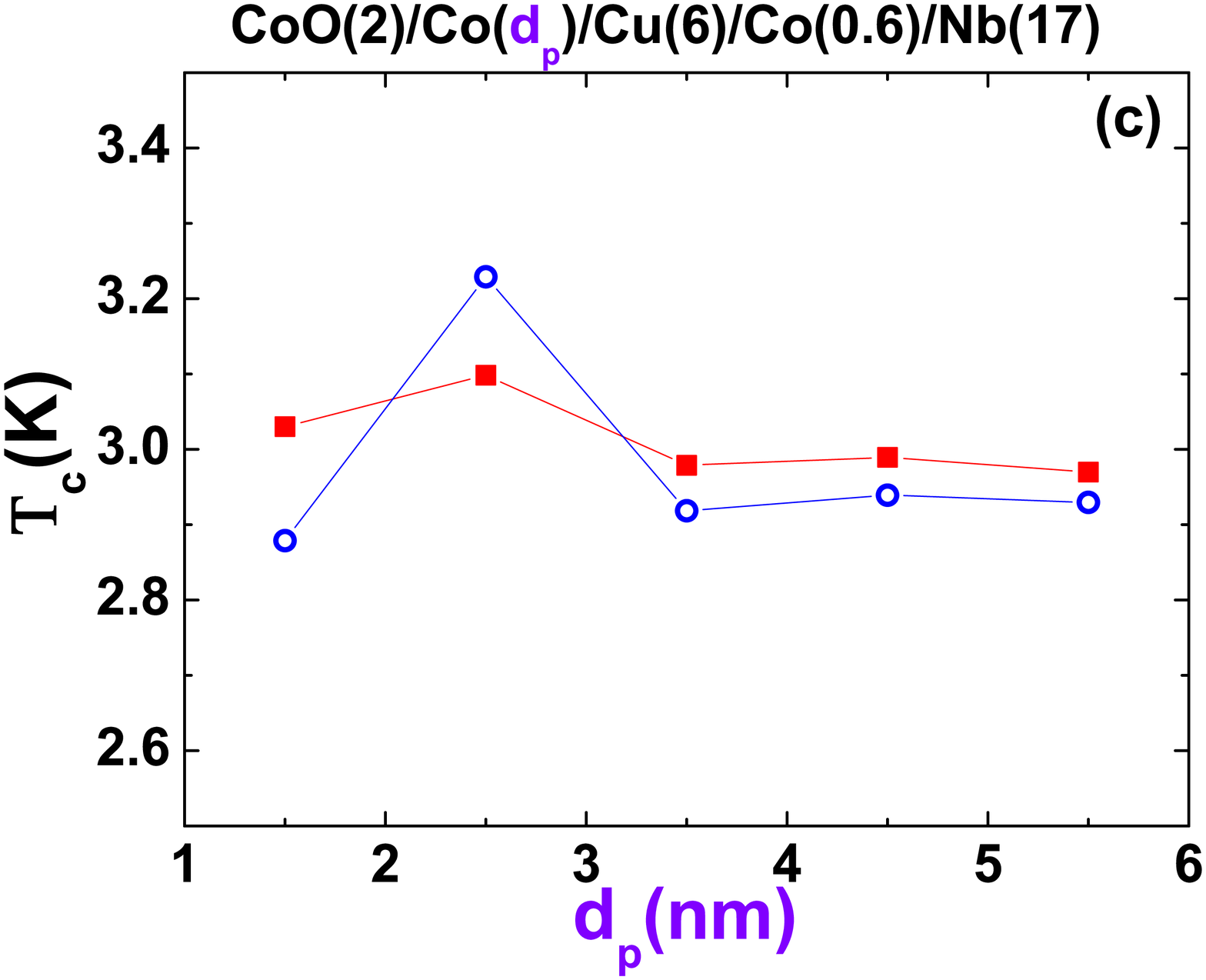}
\vspace{-2mm}
\caption{(Color online) Experimental data and theoretical fitting of $T_c$ in the P state as a function of (a) the Co free layer thickness $d_{f}$
 (with $d_n=6$ nm and $d_p=2.5$ nm), (b) the Cu normal metal layer thickness $d_n$ (with $d_p=2.5$ nm and $d_f=0.6$ nm), and (c) the 
Co pinned layer thickness $d_{p}$ (with $d_n=6$ nm and $d_f=0.6$ nm).}
\label{Tc_thicknesses}
\end{figure*}

In principle, one can obtain the superconducting transition temperatures by computing the temperature dependence of $\Delta(y)$ and identifying the critical temperature where $\Delta(y)$ vanishes. However, the property that the pair potential is vanishingly small near $T_c$ permits one to linearize the self-consistency condition, that is, to rewrite it near $T_c$ in the form
\begin{align}
\Delta_i=\sum_q J_{iq}\Delta_q, 
\end{align}
where the $\Delta_i$ are expansion coefficients in a given basis and the $J_{iq}$ are the appropriate matrix elements with respect to the same basis. To determine $T_c$, one can simply compare the largest eigenvalue, $\lambda$, of the matrix $J$ with unity at a given temperature. The system is in the superconducting state when $\lambda$ is greater than unity. More details of this efficient technique are discussed in Refs.~\cite{Zhu2010,Barsic2007}.

To analyze the correlation between the behavior of the superconducting transition temperatures and the existence of odd triplet superconducting correlations in our systems, we compute the induced triplet pairing amplitudes which we denote as $f_0$ (with $m=0$ spin projection) and $f_1$ with ($m=\pm 1$) according to the equations \cite{Halterman2007,Halterman2008}
\begin{subequations}
\label{alltripleta}
\begin{align}
f_0 (y,t) & = \frac{1}{2} \sum_n \left[ u_{n\uparrow} (y) v_{n\downarrow}(y)-
u_{n\downarrow}(y) v_{n\uparrow} (y) \right] \zeta_n(t), 
\label{f0defa} \\
f_1 (y,t) & = \frac{1}{2} \sum_n \left[ u_{n\uparrow} (y) v_{n\uparrow}(y)+
u_{n\downarrow}(y) v_{n\downarrow} (y) \right] \zeta_n(t),
\label{f1defa}
\end{align}
\end{subequations}
where $\zeta_n(t) \equiv \cos(\epsilon_n t)-i \sin(\epsilon_n t) \tanh(\epsilon_n /(2T))$. These triplet pair amplitudes are odd in time $t$ and vanish at $t=0$, in accordance with the Pauli exclusion principle. 

\section{ANALYSIS}
\vspace{-3mm}
In this subsection, we present our theoretical analysis and compare the theoretical results with the experimental data. To find the theoretical $T_c$, we adopted the linearization method as discussed in Sec.~\ref{methods}. The fitting process is rather time-consuming since for every parameter set, one must evaluate $T_c$ numerically as a function of the misalignment angle $\alpha$, making a least-squares fit unfeasible. The same situation occurs in Refs.~\cite{Zhu2010,Chiodi2013}. As in those works, we search within plausible regions  of parameter space, and display here results of the best fit that we have found, which is not necessarily the best possible fit. There are a number of parameters at one's disposal and, when computing the theoretical values of $T_c$, we first have to keep the number of fitting parameters as small as possible. All of the relevant physical parameters that are related to the properties of the materials involved, such as the exchange field, and the  effective superconducting coherence length, are required to be the same for all of the different samples when performing the fitting. However, for parameters that are affected by the fabrication processes such as the interfacial barrier strength, one can reasonably assume, as we do,  that their values are somewhat different from sample to sample. We do find that the variation is small between different samples in each series. For the material parameters we have found that the best value of the effective  Fermi wave vector is $k_F=1\AA^{-1}$ and the effective superconducting coherence length $\xi_0=11.5 \,\textrm{nm}$. For the  dimensionless exchange field $I\equiv h_0/E_F$ (normalized to Fermi energy), we have used, for \rm{Co}, $I=0.145$, which is consistent with previous work \cite{Halterman2008} ($I=1$ corresponds to the half-metallic limit). For the superconducting transition temperature for a putative pure superconducting sample of the same quality as the material in the layers, we have used $T_c^0=4.5$ K. This is the same value previously found \cite{Zhu2010}. It  is of course lower than the true bulk transition temperature of \rm{Nb} but even for pure thin films a decrease in $T_c$ is to be expected \cite{wolf1975}.  All of these parameters are kept invariant across all of the different samples, as mentioned earlier. Only the three interfacial barrier strengths are treated as adjustable from sample to sample during the fitting process. We assume, however, that the barrier strength is the same on both sides of the normal metal layer while that between the free ferromagnetic layer and the superconductor are weaker. For each series, the barrier varies somewhat from batch to batch. 

\begin{figure*}[htb] 
\centering
\includegraphics[width=0.73244\columnwidth]{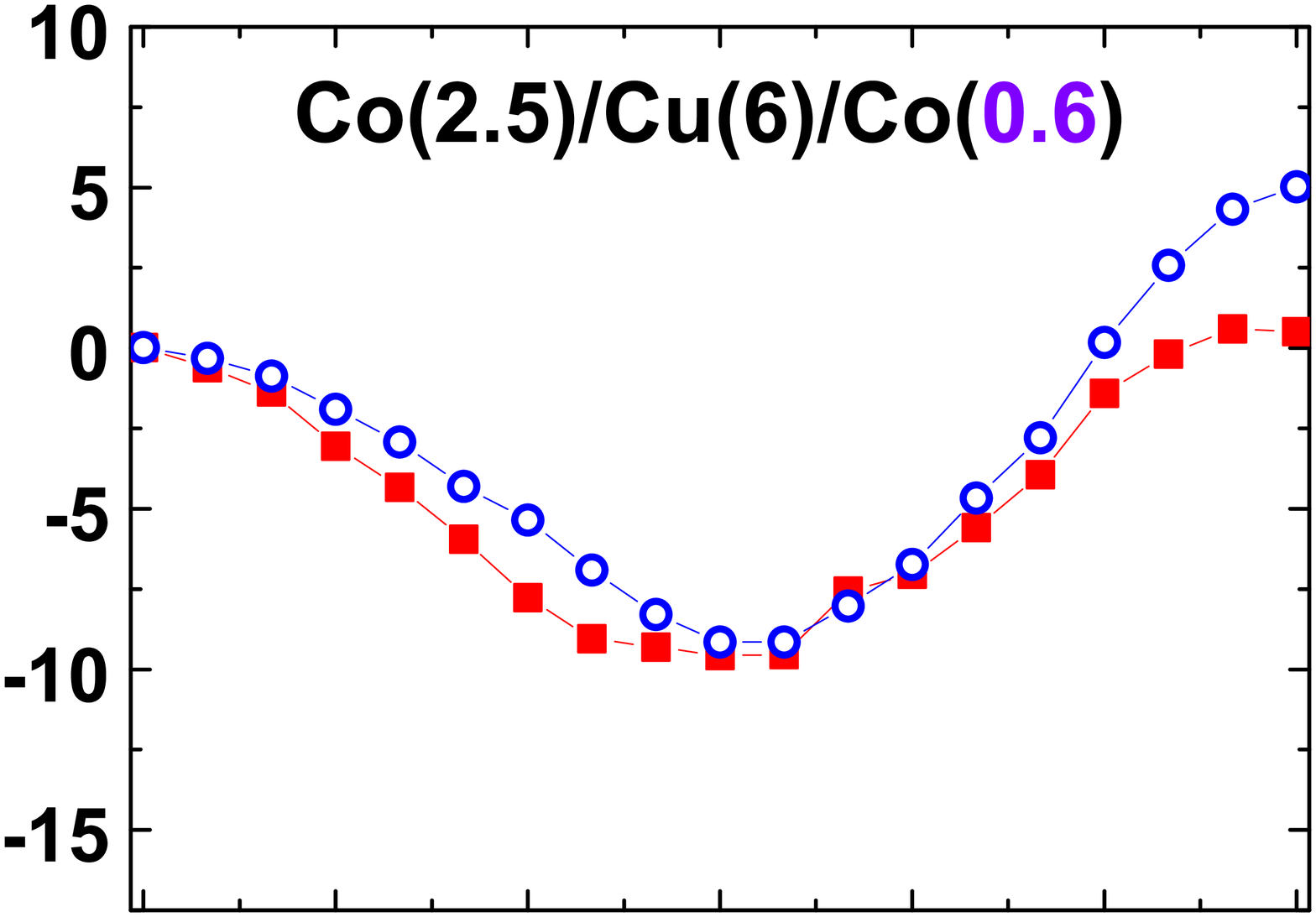}
\includegraphics[width=0.63378\columnwidth]{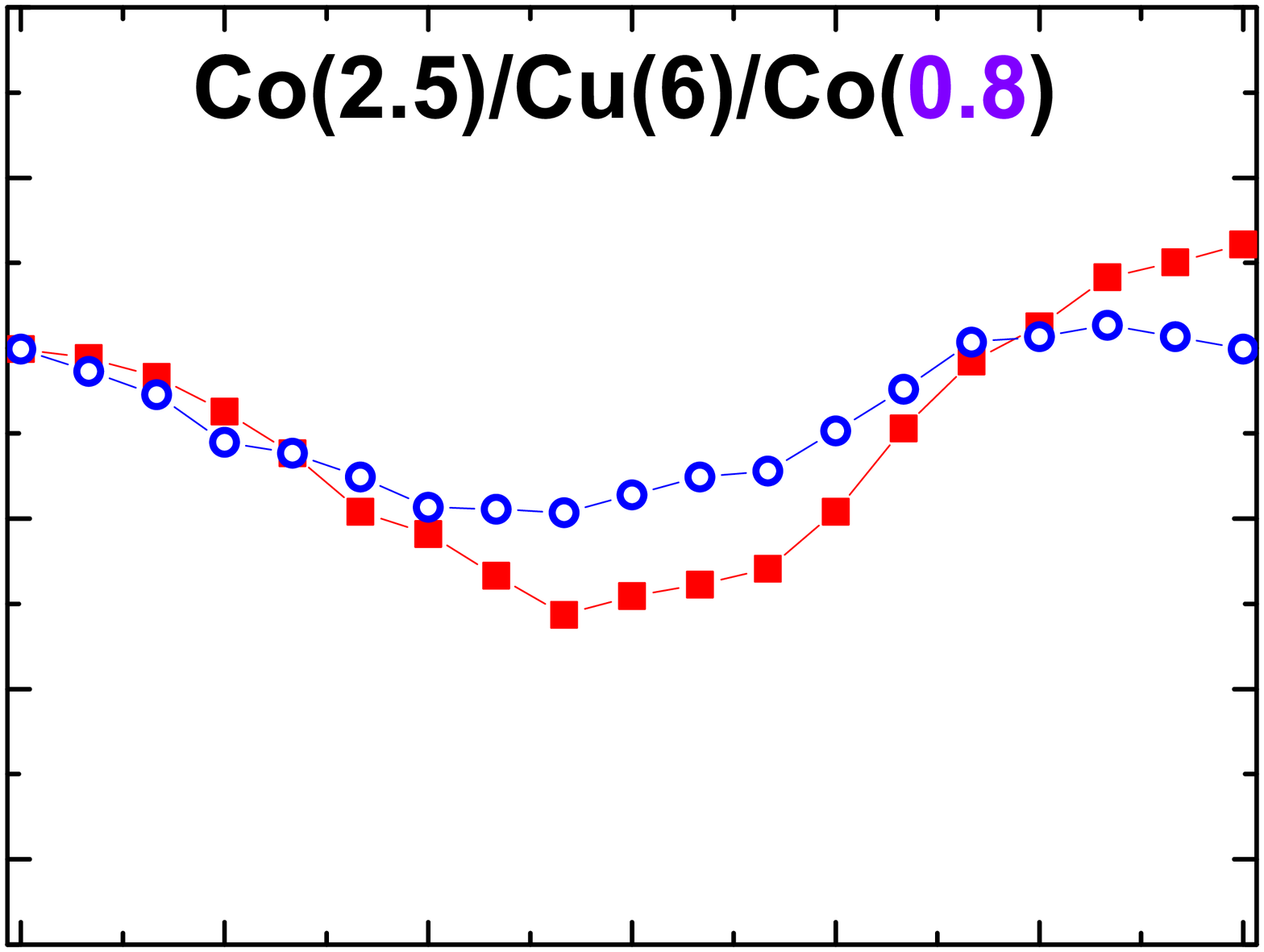}
\includegraphics[width=0.63378\columnwidth]{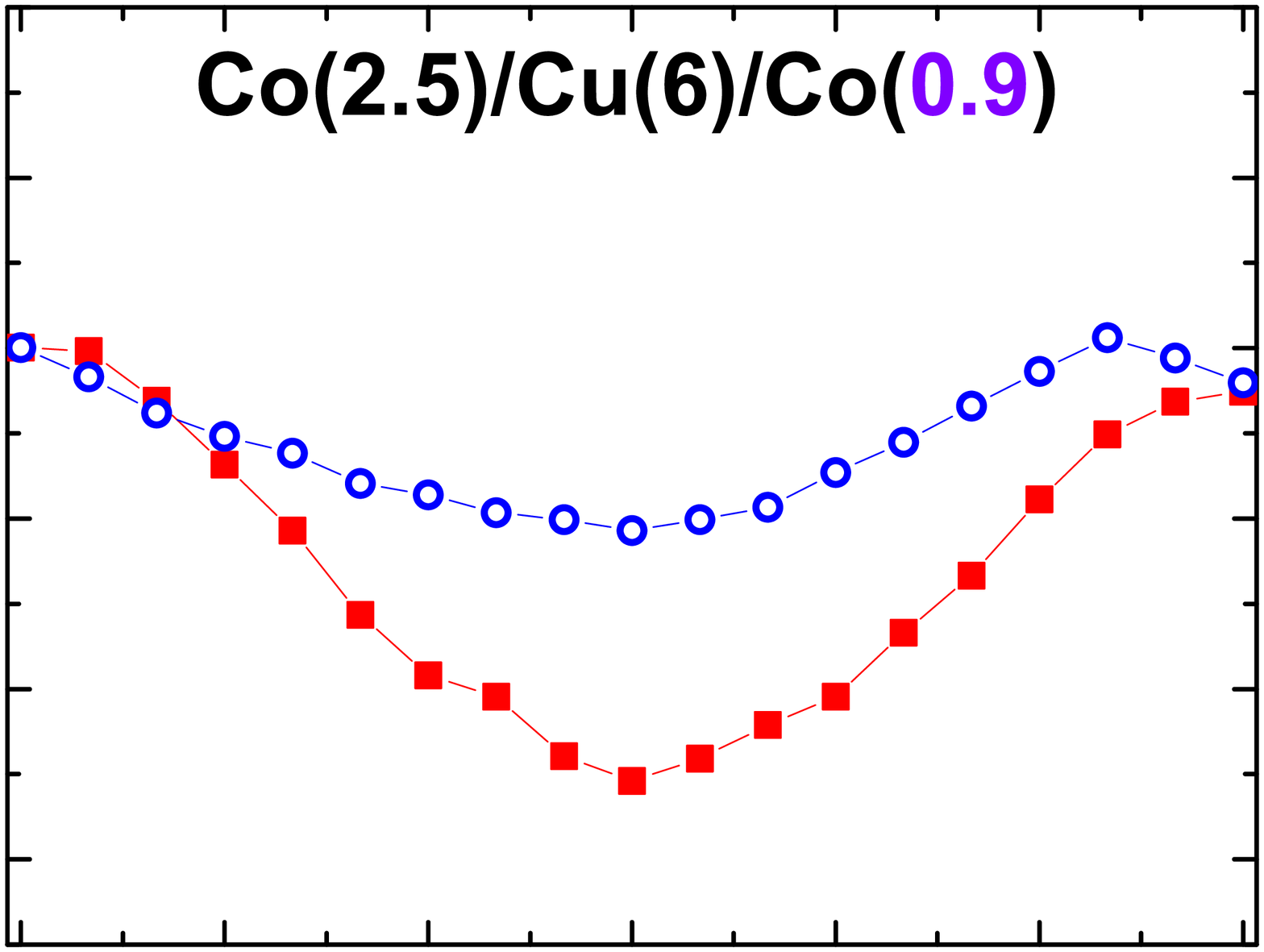}\\
\includegraphics[width=0.73244\columnwidth]{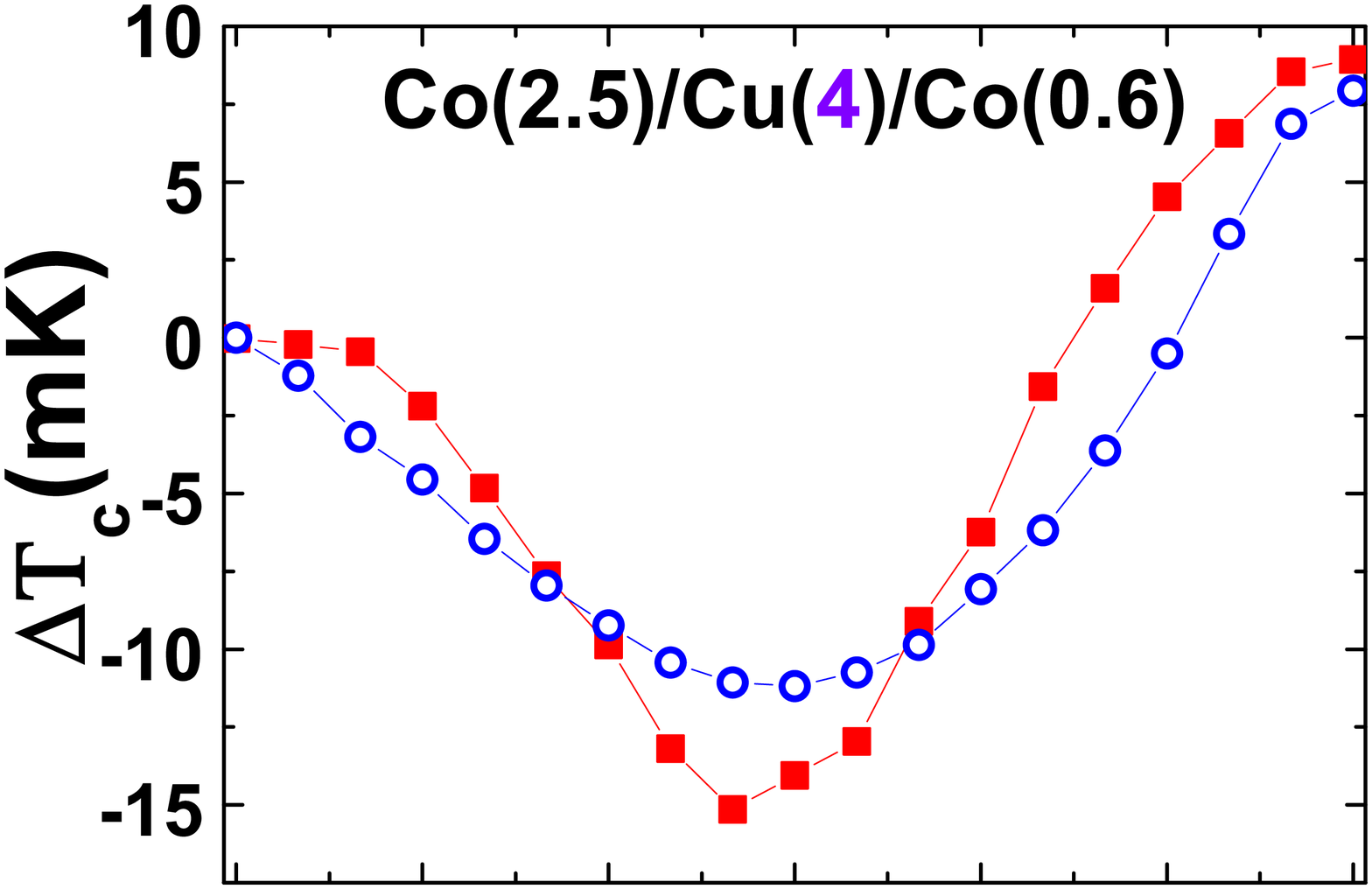}
\includegraphics[width=0.63378\columnwidth]{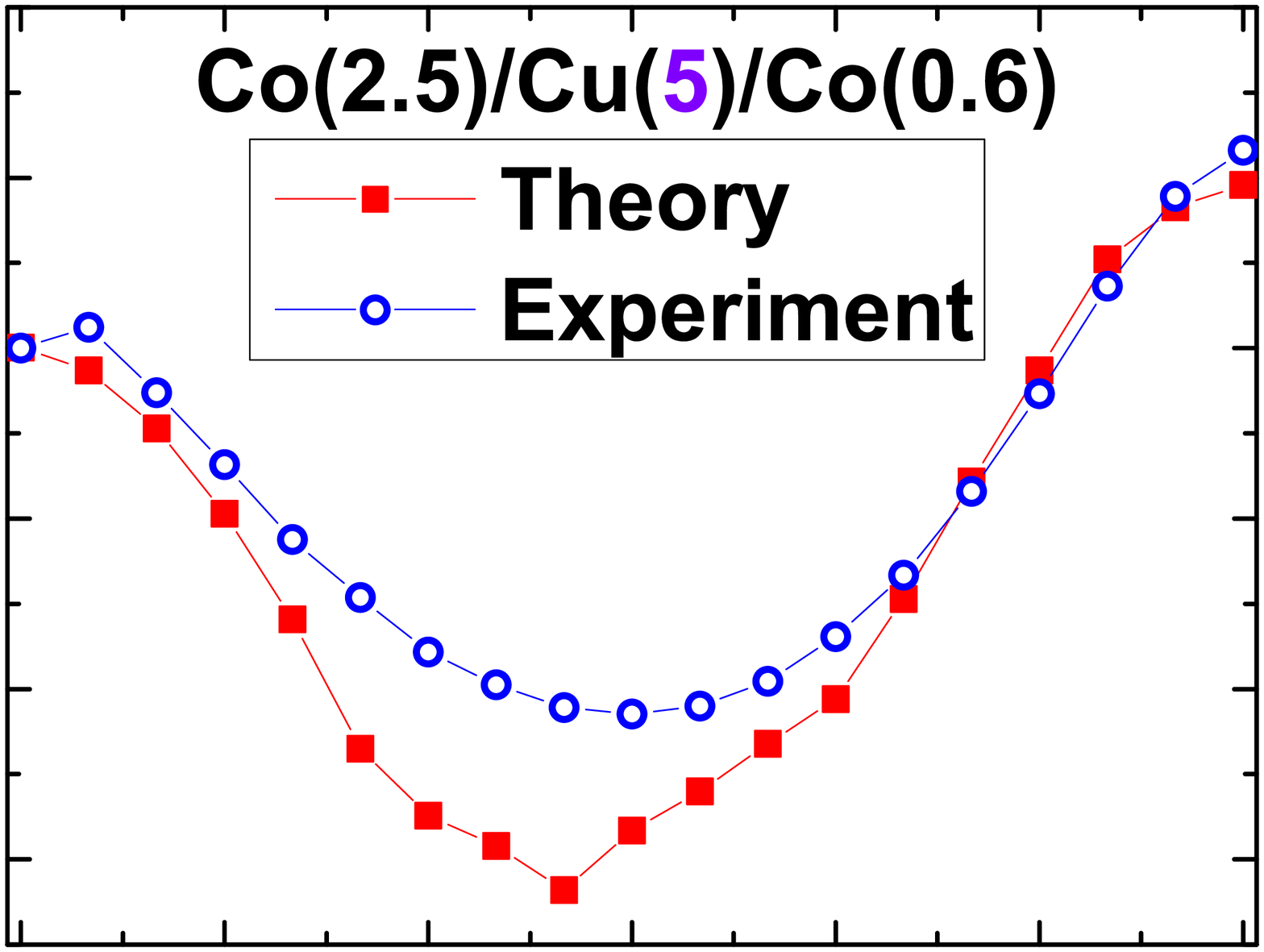}
\includegraphics[width=0.63378\columnwidth]{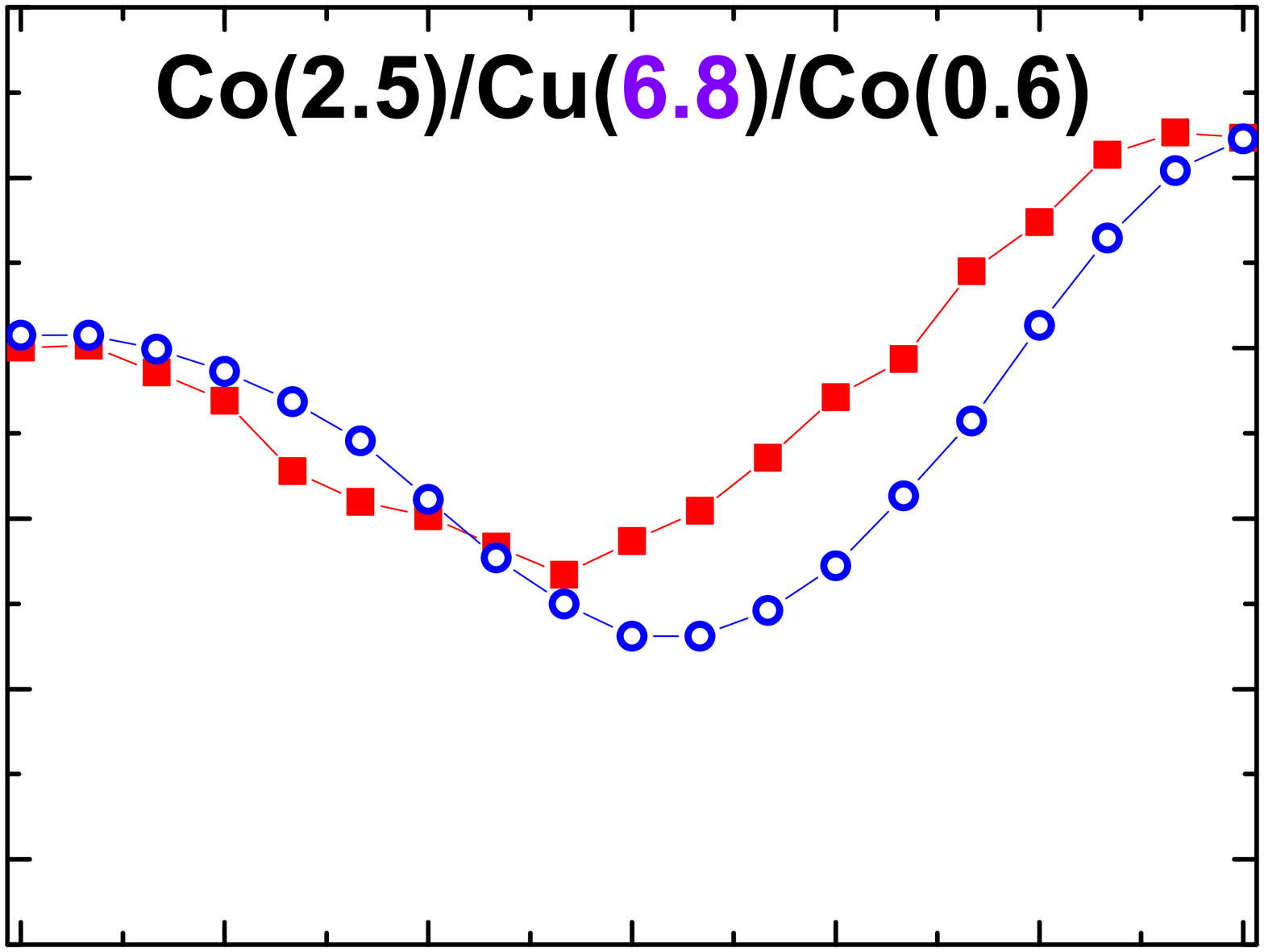}\\
\includegraphics[width=0.73244\columnwidth]{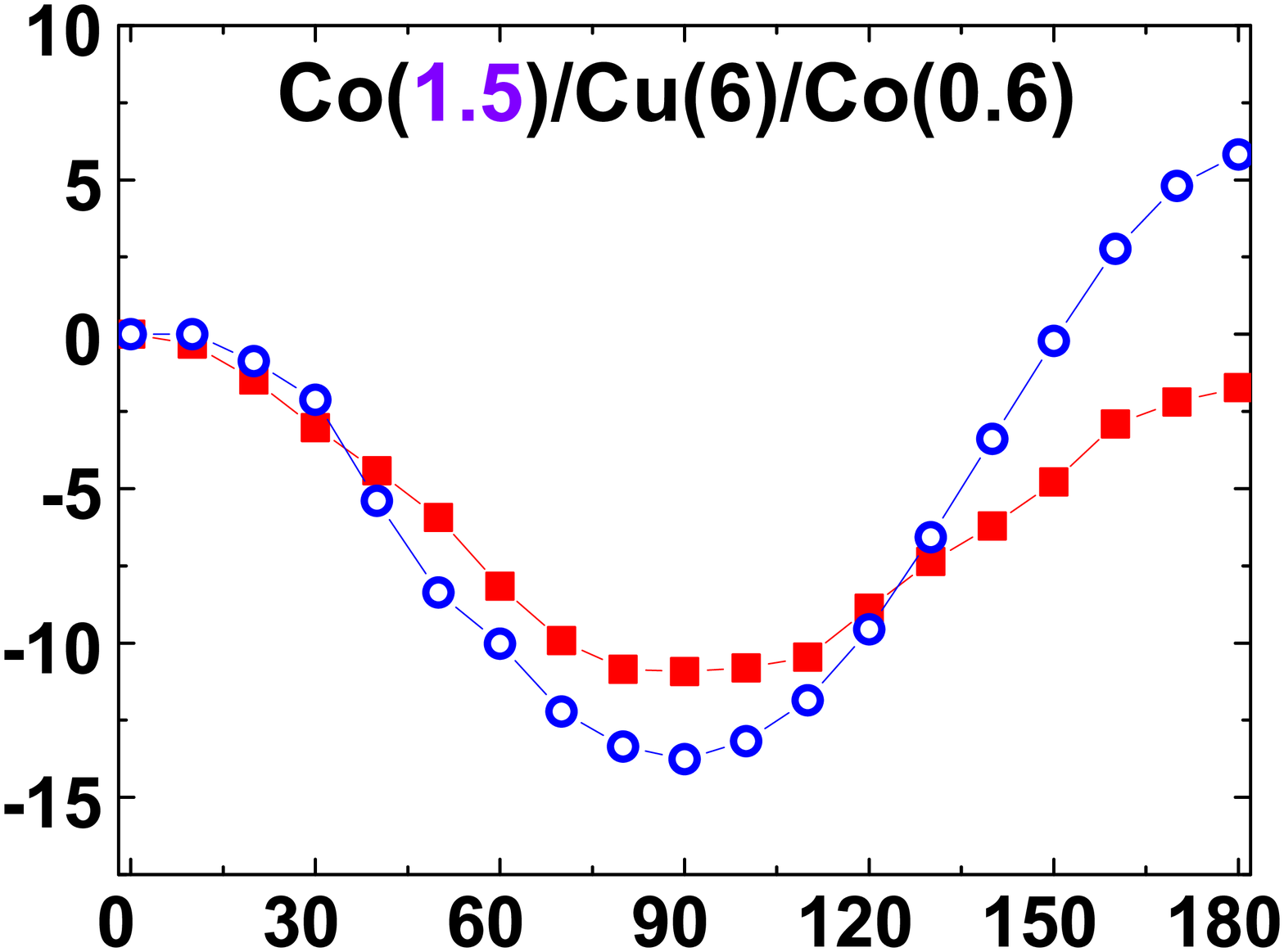}
\includegraphics[width=0.63378\columnwidth]{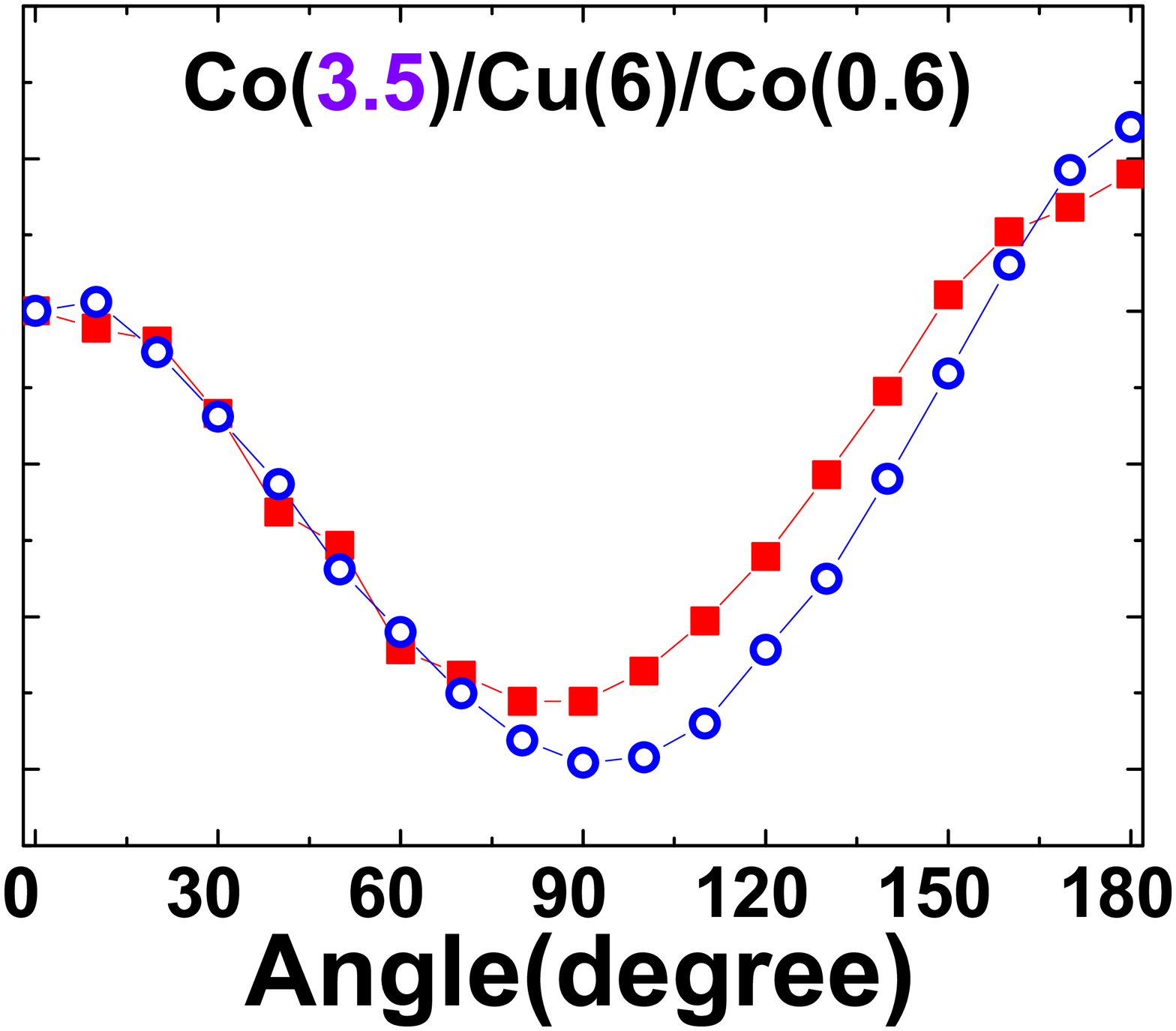}
\includegraphics[width=0.63378\columnwidth]{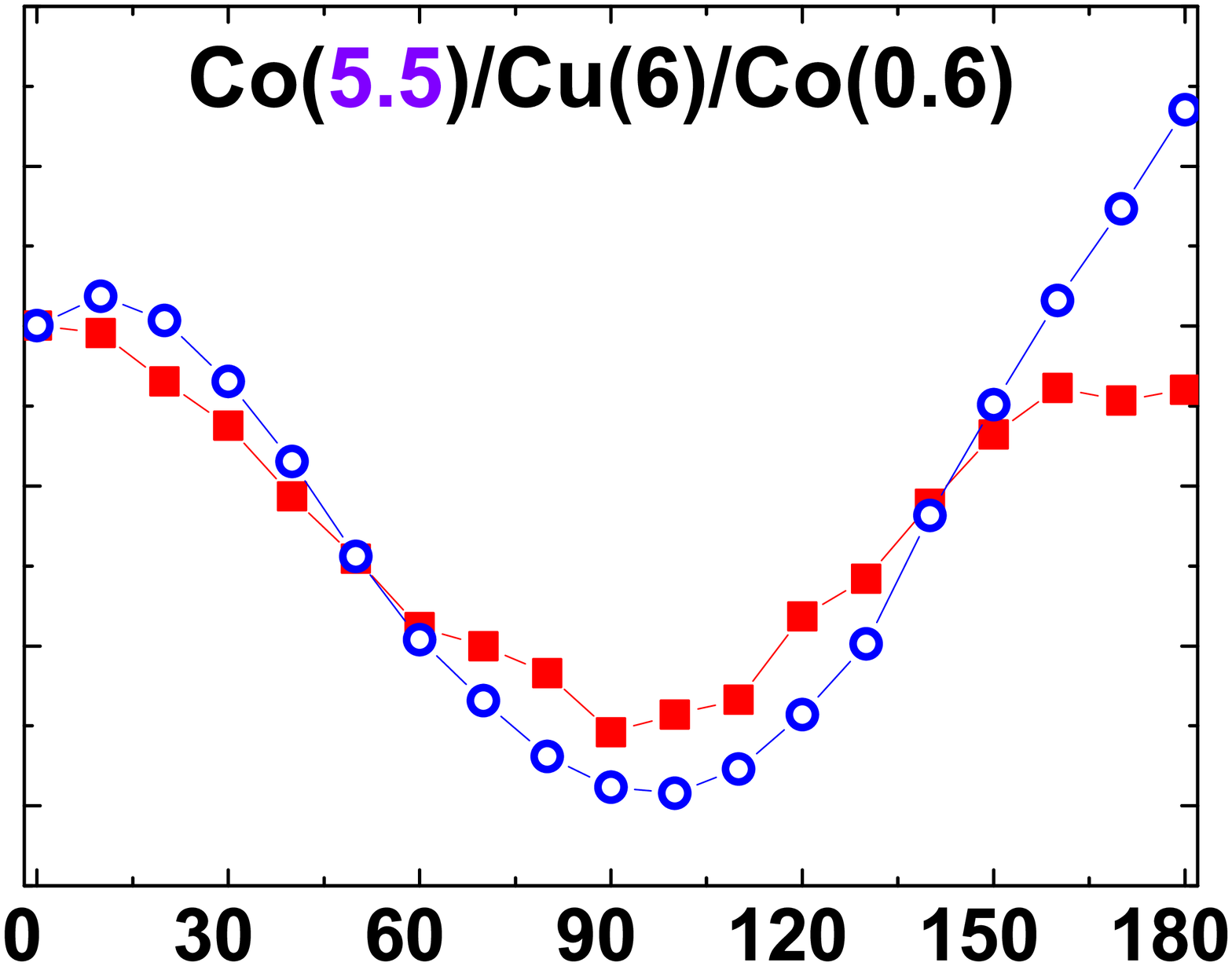}
\vspace{-2mm}
\caption{(Color online) Experiment and theory comparisons of $\Delta T_c$ [defined as $\Delta T_c(\alpha)\equiv T_c(\alpha)-T_c(0)$] as a function of relative magnetization angle are shown for the three batches of samples. Top row: Three different free layer thicknesses, $d_{f}=$ 0.6 nm, 0.8 nm, 0.9 nm, and with $d_{p}=$ 2.5 nm, $d_n=$ 6 nm. Middle row: Three different nonmagnetic layer thicknesses: $d_n=$ 4 nm, 5 nm,  6.8 nm, and with $d_{f}=$ 0.6 nm, $d_{p}=$ 2.5 nm. Bottom row: Three different pinned layer thicknesses:  $d_{p}=$ 1.5 nm, 3.5 nm, 5.5 nm, and with $d_{f}=$ 0.6 nm, $d_n=$ 6 nm. } 
\label{All}
\end{figure*}

They are found to be as follows: $H_{B1}=0.2$, and both $H_{B2}$ and $H_{B3}$ vary from $0.64$ to $0.7$ for different batches in the $d_{f}$ series. For the $d_{p}$ series, we have $H_{B1}=0.15$, $0.53<H_{B2}$, and $H_{B3}<0.58$. The $d_n$ series have $H_{B1}$ ranges from $0.3$ to $0.45$ and $H_{B1}=H_{B2}=0.62$.  The thicknesses of the different layers are taken of course from their experimental values. As in Ref.~\cite{Chiodi2013} we find a thin magnetic ``dead layer'' between the normal metal and the free ferromagnetic layer of a small thickness in the range $0.27 \,\textrm {nm} \sim 0.35 \,\textrm {nm}$.

We now compare the experimental and theoretical values of $T_c$ as a function of layer thicknesses and angle $\alpha$ for three different batches of samples: in one we vary $d_{f}$, in the second, $d_{n}$, and in the last, $d_{p}$. First, in Fig. \ref{Tc_thicknesses}, we present comparisons between experiment and theory, for the $T_c$ results in the parallel state ($\alpha=0$) as a function of thickness for the three different series mentioned above. In all three series, the experimental and theoretical $T_c$ are in very good agreement with each other. For the $d_{f}$ series, one should notice that both experimental and theoretical $T_c$ are very sensitive to the thicknesses of the free layers. When the thickness of the free ferromagnetic layer is increased, $T_c$ decreases nonmonotonically by almost $50 \%$. However, the $d_n$ and $d_{p}$ series do not show the same sensitivity, even though the ranges of thicknesses for these two series are much larger compared to that of the $d_{f}$ series. This  lower sensitivity is physically reasonable for the following reason: because of the presence of ferromagnets, we find that the magnitude of the singlet pairing amplitude decreases very fast beyond the boundary, in non-$\rm S$ regions away from the $\rm F/S$ interface. The exchange field reduces the proximity effect. Therefore, the size effects from the thicknesses of normal metal layers and pinned ferromagnetic layers are less. We also observe the trend that both theoretical and experimental $T_c$ are often found to be a nonmonotonic function of the thicknesses of the $\rm F$ layers. In fact, except for the experimental $T_c$ for $d_{f}$ series, which does not show a clear oscillatory behavior, all other series clearly exhibit the nonmonotonicity of $T_c$. Oscillatory behavior of transition temperatures as one varies the thickness is standard in hybrid ${\rm S/F}$ heterostructures due to the oscillatory character of the pair amplitude \cite{Demler1997} itself. The reason for the exception found might be that the data points are too widely spaced. This nonmonotonic behavior has been noted in past works \cite{Radovic1991,Jiang1995} and is often found \cite{Wu2012} in FFS trilayers.

\begin{figure*}[htb]
\centering 
\includegraphics[width=0.66\columnwidth]{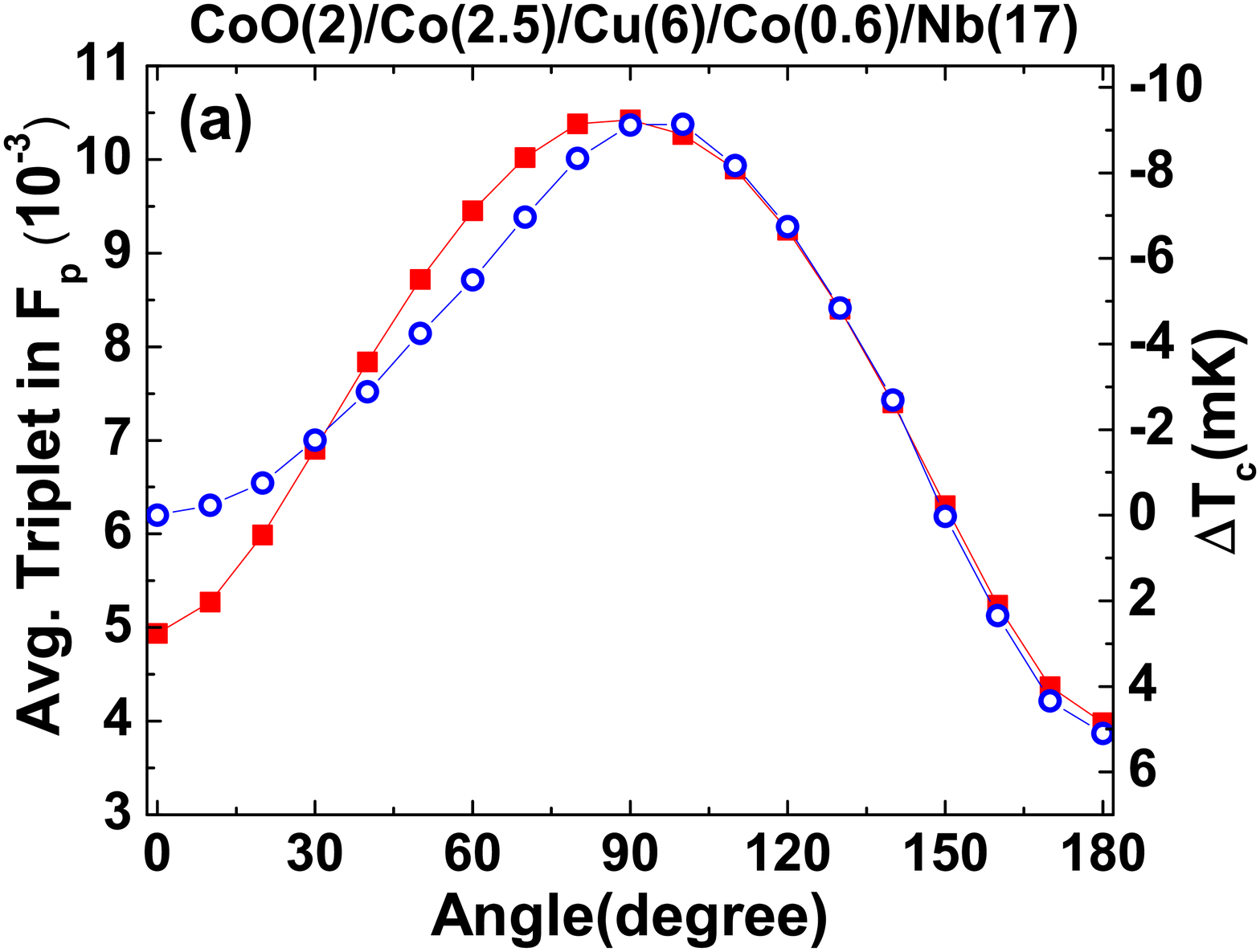}
\includegraphics[width=0.66\columnwidth]{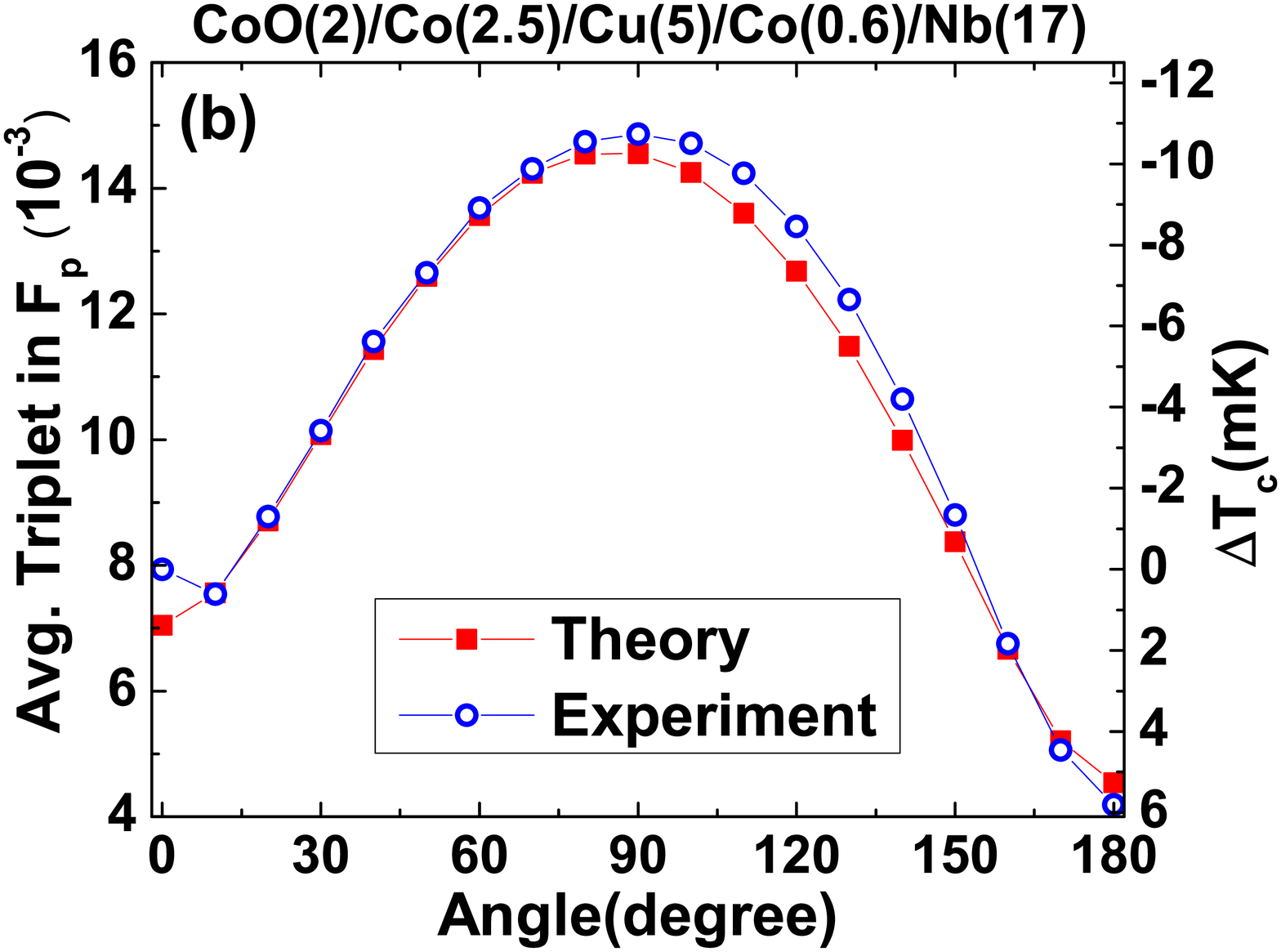}
\includegraphics[width=0.66\columnwidth]{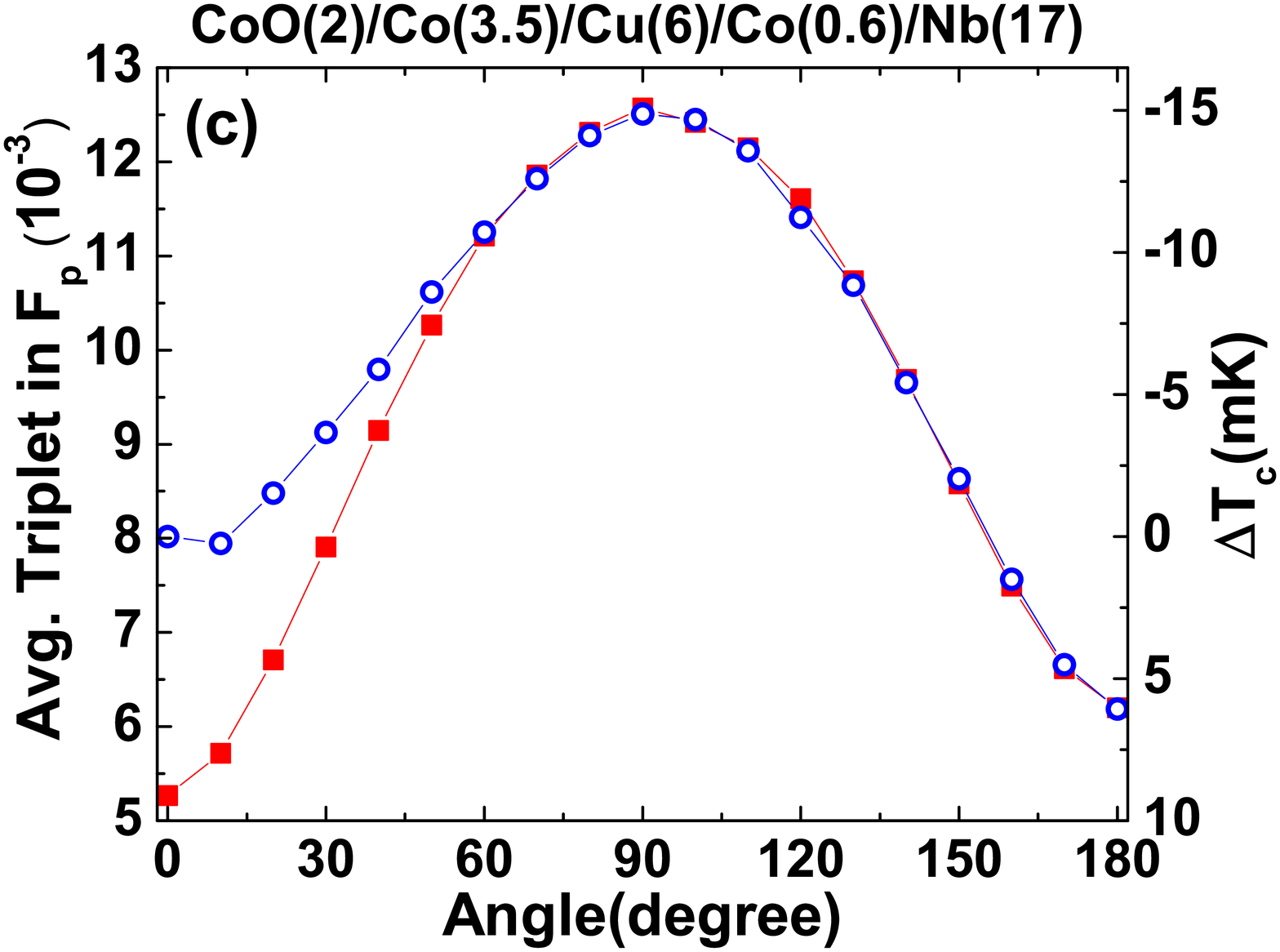}
\vspace{-2mm}
\caption{(Color online) Average triplet amplitudes in the pinned ferromagnet layer as a function of relative magnetization angle. The quantity plotted is the average of $F_t(y,t)$ [Eq.~(\ref{tripletq})] in this region, at $\omega_D t=4$. The quantity $\Delta T_c$ is also shown (right scale). Red squares are the theoretical triplet amplitudes (left scale) and the blue circles are the experimental $\Delta T_c$ (right inverted scale) as a  function of angle. 
The $\Delta T_c$ data correspond to one set chosen from each batch of samples in Fig.~\ref{All}. (a) From the $d_{f}$ series, (b) from the $d_n$ series, (c) from the $d_{p}$ series.}
\label{Triplet_DTc_A_All}
\end{figure*}

In Fig.~\ref{All}, we present a detailed comparison of theoretical and experimental results for $\Delta T_c$ as a function of angle $\alpha$ between the magnetizations in the free and pinned layers for the $d_{f}$, $d_{n}$, and $d_{p}$ series. Each panel in the first row in Fig.~\ref{All} represents different samples for $d_{f}$ series. Results for the $d_{n}$ and $d_{p}$ series are  plotted in the second and third row, respectively. One can clearly see that the  behavior of the highly nonmonotonic angular dependencies of the theoretical results presented here describe very well  the experimental results, not only qualitatively but also quantitatively: the magnitudes of the experimental  and theoretical results for $\Delta T_c$ are comparable; both experimental and theoretical results indicate the switching effects are in about the $25$ mK range. It is well worth recalling than in another recent work \cite{Zhu2010} results for the magnitude of this quantity differed  by more than one order of magnitude. In contrast, here,  taking into account the existence of numerical and experimental uncertainties (the former we estimate at $\sim 1.5$ mK), we find theory and experiment in very good agreement. This great improvement over Ref.~\cite{Zhu2010} follows from the more careful treatment of the interface barriers from sample to sample and a much more extensive search in parameter space. For the $d_{f}$ series, we see that the switching range for both experimental and theoretical $T_c(\alpha)$ varies nonmonotonically when $d_{f}$ is increased. This occurs for the same reason already mentioned in the discussion of Fig.~\ref{Tc_thicknesses}: the behavior of $T_c(\alpha)$ is very sensitive to the inner ferromagnetic layer thicknesses due to the proximity effect. Similarly, we observe that the switching ranges are less sensitive to the thickness of the outer ferromagnetic layer (see in the $d_{p}$ series) and also to the normal metal layer thickness in the $d_n$ series.

We now turn to the role that induced triplet correlations in the nonmonotonic behavior of $T_c(\alpha)$. This has been the subject of recent theoretical interest \cite{golu,karmin,Wu2012} but little has been done on quantitatively comparing theory and experiment. To examine this question in a quantitative way, we have computed the induced odd triplet pairing correlations. These correlations (as well of course as the ordinary singlet correlations) can be self-consistently calculated using the methods previously described. As noted in  Sec.~\ref{methods}, with the presence of nonhomogeneous magnetization the triplet pair amplitudes in general can be induced when $t\neq 0$. We present our study in terms of the quantity
\begin {equation}
\label{tripletq}
F_t(y,t) \equiv \sqrt{{\lvert f_0(y,t) \rvert}^2+{\lvert f_1(y,t) \rvert}^2},
\end{equation}
where the quantities involved are defined in Eq.~(\ref{alltripleta}). This quantity accounts for both triplet components, the equal spin and opposite spin triplet correlations. The reason to use this quantity is that via Eq.~(\ref{rotation}), one can easily show that, when the spin quantization axis is rotated by an angle $\theta$, the rotated triplet pair amplitudes $\tilde{f_0}$ and $\tilde{f_1}$ after the transformation are related from the original $f_0$ and $f_1$ by
\begin{subequations}
\label{rottripleta}
\begin{align}
\tilde{f_0}(y,t) & = \cos(\theta)f_0(y,t)-\sin(\theta)f_1(y,t),
\label{f0rot} \\
\tilde{f_1}(y,t) & = \sin(\theta)f_0(y,t)+\cos(\theta)f_1(y,t).
\label{f1rot}
\end{align}
\end{subequations}
Therefore the quantity $F_t(y,t)$ that we focus on obviates any ambiguity issues related to the existence of generally non-collinear ``natural'' axes  of quantization in the system. 

We have computed this quantity as a function of position and $\alpha$. It turns out to be particularly useful to focus on the average value of $F_t(y,t)$ in the pinned layer ${\rm F_p}$. We normalize this averaged quantity, computed in the low-$T$ limit, to the value of the singlet pair amplitude in the bulk ${\rm S}$. This normalized averaged quantity is plotted as a function of $\alpha$ in Fig.~\ref{Triplet_DTc_A_All} (left vertical scale) at a dimensionless characteristic time $\omega_D t=4.0$. This time value is unimportant, provided it be nonzero, of course. In the three panels, an example taken from each of the series is displayed, as explained in the caption. One can observe that the maxima of this average $F_t$ occur when $\alpha=\pi/2$ and its minima are at either $\alpha=0$ or $\alpha=\pi$. In the same figure (right vertical scale) the experimental values of $\Delta T_c(\alpha)$, for the same cases, which have minima near $\pi/2$, are plotted in an inverted scale. The agreement is truly striking. The anticorrelation can be easily understood: the magnitude of the low-$T$ singlet pair amplitudes is of course positively correlated to $T_c$. Here the fact that triplet pair amplitudes are anticorrelated to $T_c$ (or to the singlet amplitudes) indicates a singlet-triplet conversion process: when more singlet superconductivity leaks into the ferromagnet side, $T_c$ is suppressed and triplet superconductivity is enhanced. The average magnitude of the triplet pair amplitudes in the free and normal layer regions is only weakly dependent on $\alpha$: Of importance is the propagation of triplet pairs throughout the entire system, generated by the symmetry-breaking interfaces and magnetic inhomogeneity created from the two misaligned ferromagnets. This clearly demonstrates a singlet to triplet process which is related to the nonmonotonicity of the transition temperature. 

\section{Conclusion} 

In conclusion, we made measurements of the superconducting transition temperature $T_c$ in CoO/Co/Cu/Co/Nb multilayers in a spin valve structure. $T_c$ was measured both as a function of the in-plane angle between the Co magnetic moments and of the thicknesses of the Co/Cu/Co spin valve layers. We found that $T_c$ is a nonmonotonic function of the angle, with a minimum near orthogonal orientations of the magnetic moments of the two Co layers. The behavior of $T_c$ as a function of these variables was quantitatively described by an efficient microscopic method that is based on a linearization of the self-consistent Bogoliubov--de Gennes equations. We have shown that the nonmonotonic behavior of $T_c(\alpha)$ is correlated with the formation of long-range triplet pairs.

\section{Acknowledgements}

This work was supported by IARPA under Grant No. N66001-12-1-2023. A.A.J acknowledges support from US-Chile Equal Opportunities Scholarship from FULBRIGHT-CONICYT. C.-T.W acknowledges support from a Dissertation Fellowship from the University of Minnesota Graduate School.

\end{document}